\documentclass[%
reprint,
superscriptaddress,
 amsmath,amssymb,
 aps,
]{revtex4-2}
\usepackage{amsmath}
\usepackage{amssymb}
\usepackage{graphicx}
\usepackage{dcolumn}
\usepackage{bm}
\usepackage{hyperref}
\usepackage{fixfoot}
\usepackage{color}
\usepackage{xcolor}
\usepackage{subcaption}
\usepackage{array}

\begin{document}

\title{
Empirical validation of the polarization transition in a double-random field model of elections
}%

\author{Jan Korbel}
\affiliation{Complexity Science Hub, Metternichgasse 8, 1030, Vienna, Austria}

\author{Remah Dahdoul}
\affiliation{Complexity Science Hub, Metternichgasse 8, 1030, Vienna, Austria}

\author{Stefan Thurner}
\email{stefan.thurner@meduniwien.ac.at}
\affiliation{Section for the Science of Complex Systems, Center for Medical Data Science, Medical University of Vienna, Spitalgasse 23, A-1090 Vienna, Austria}
\affiliation{Complexity Science Hub, Metternichgasse 8, 1030, Vienna, Austria}
\affiliation{Santa Fe Institute, 1399 Hyde Park Road, Santa Fe, New Mexico 87501, USA}


\begin{abstract}
We model bipartisan elections where voters are exposed to two forces: local homophilic interactions and external influence from two political campaigns. The model is mathematically equivalent to the random field Ising model with a bimodal field. When both parties exceed a critical campaign spending, the system undergoes a phase transition to a highly polarized state where homophilic influence becomes negligible, and election outcomes mirror the proportion of voters aligned with each campaign, independent of total spending. The model predicts a hysteresis region, where the election results are not determined by campaign spending but by incumbency. Calibrating the model with historical data from US House elections between 1980 and 2020, we find the critical campaign spending to be $\sim 1.8$ million USD. Campaigns exceeding critical expenditures increased in 2018 and 2020, suggesting a boost in political polarization.
\end{abstract}

\maketitle

Classical statistical physics provides a framework for understanding collective phenomena. It typically assumes weakly interacting constituents, governed by a single, time-invariant force, which allows computation of phase diagrams and identification of critical points where system-wide change occurs. This framework has long been applied to social phenomena, including financial markets \cite{Mantegna1995,Gabaix2003,Bouchaud2023}, pedestrian and crowd dynamics \cite{Helbing1995,Karamouzas2014}, anthropology \cite{Lansing2017,Gandica2021}, judicial systems \cite{Lee2015}, and opinion formation \cite{Galam1999,Neirotti2024,Klimek_2008,Castellano2009}. In particular, it underpins opinion dynamics models such as bounded confidence \cite{Hegselmann2002,Deffuant2000}, social validation \cite{Sznajd2000}, and cultural dissemination \cite{Axelrod1997}.

In voting behavior, one of the most influential physics-inspired models is the \emph{voter model} \cite{Liggett1999}, where agents adopt neighbors’ opinions on lattices or heterogeneous networks \cite{Sood2005}. Extensions include stochastic noise \cite{Granovsky1995}, committed agents or ``zealots''~\cite{Mobilia2003}, and the interplay of micro- and macrodynamics \cite{Stark2008}. It has been shown that the voter model reproduces statistical features of US presidential elections \cite{Garcia2014}; see also \cite{Redner2019} for a review.

A more recent class of opinion dynamics models builds on two sociological principles: homophily—the tendency to associate with similar others \cite{McPherson2001}—and social balance—the reduction of cognitive dissonance in triads \cite{Heider1946, Marvel2009}. Extending the voter model, these approaches capture more realistic interactions. Their combined effects have been studied jointly \cite{Pham2020, Gorski2020} and integrated into a unified framework \cite{Pham2022}, applied to group formation \cite{Korbel2023}, and recently validated experimentally \cite{Galesic2025}. Many such models, including voter- and homophily-based ones, are inspired by the Ising model, long a versatile tool in various interdisciplinary contexts \cite{Macy2024}, especially in opinion dynamics \cite{Mullick2025,starnini2025,caldarelli2025}. In elections, it explained the universal scaling of vote distributions in proportional systems \cite{Fortunato2007}, predicted margins of victory from turnout \cite{Pal2025}, and modeled temporal effects through an external field \cite{Tiwari2021}.

A central challenge in complex adaptive systems such as societies is the variety of time-dependent interactions, co-evolution, herding, and anticipation of others’ actions \cite{thurner2018}. While some aspects are easy to model, calibration and validation remain difficult, requiring great care to ensure testable models. Here, we address opinion formation with multiple interaction types, as in political elections. Voters exchange views within social networks of family, friends, and colleagues, while also following political campaigns—typically only those of their preferred party. We model these two processes through Ising interactions for homophily and a bimodal random field for campaign influence. Together, they form a \emph{Random Field Ising Model} (RFIM), extending spin-spin interactions with site-dependent random external fields. An equivalent mean-field description can be derived from a master-equation of stochastic opinion switching with preference and adaptation terms, as originally introduced in the context of sociodynamics \cite{weidlich1991physics,weidlich2012concepts}.

\begin{figure*}[h]
    \includegraphics[width=0.45\linewidth]{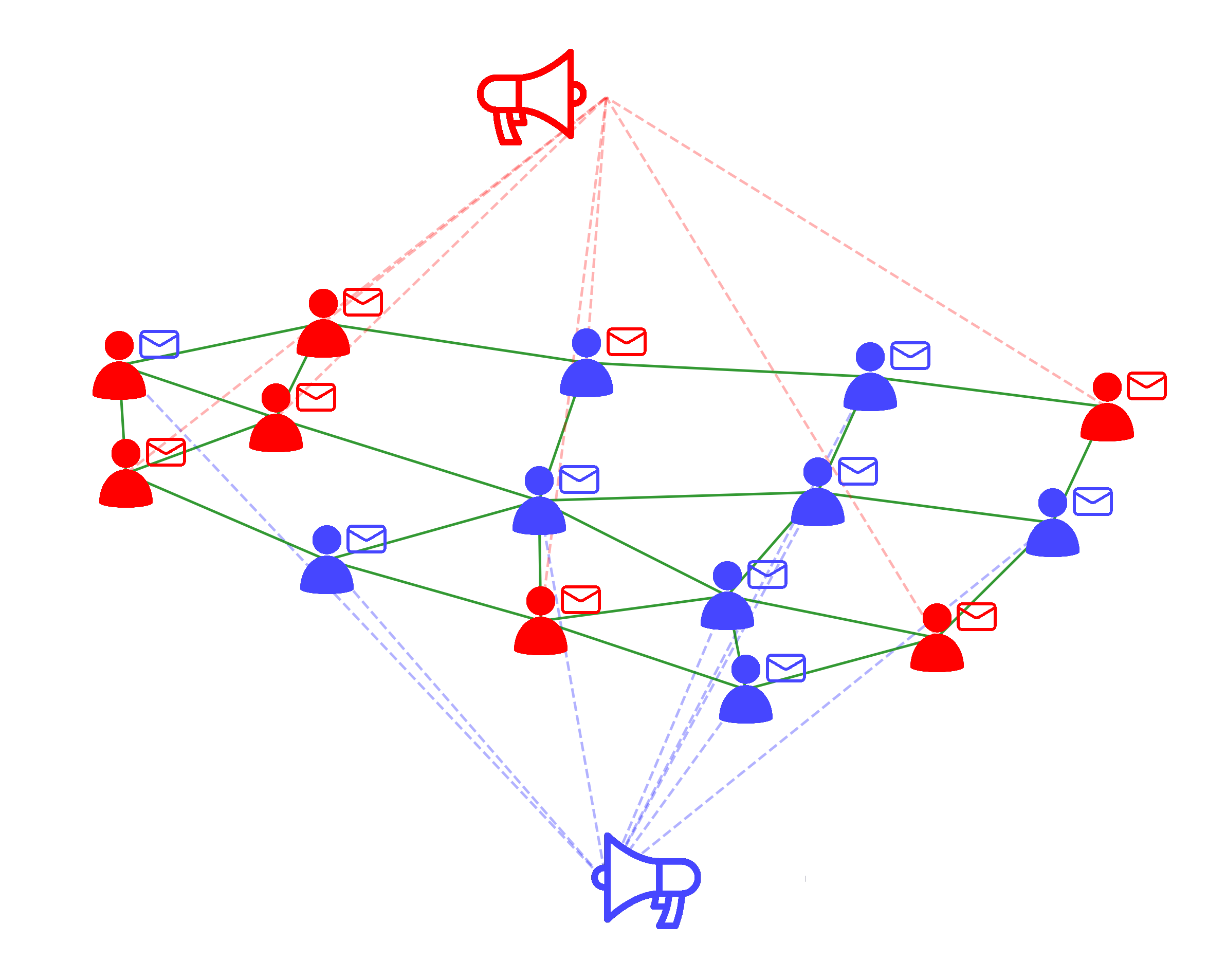}
    \caption{\textbf{Illustration of the model of voters influenced by homophily and election campaign}. Every individual has a binary opinion, expressing their voting preference. Everyone is following one of the political campaigns, while also being influenced by their local social environment (friends) in homophilic interactions with the neighbors in the social network.}
    \label{fig:syst}
\end{figure*}

Despite its simplicity, the RFIM captures rich behavior such as quenched disorder and complex phase diagrams \cite{Imry1975, Bricmont1987, Fytas2018}. Variants with bimodal random fields, where the field takes two values \cite{Hartmann1999, Sinova2001, Fytas2008}, exhibit tri-critical points marking the transition from second- to first-order phase changes \cite{Aharony1987,rfim2010}. In the context of an election campaign, the RFIM represents a bipartisan electorate where each voter holds a binary preference. Voters occupy a social (friendship) network and are randomly assigned one of two field values: the sign encodes campaign affiliation, while the magnitude reflects campaign strength, with spending serving as a proxy. A schematic illustration is shown in End Matter, Fig.~\ref{fig:syst}. RFIM approaches have long been used in sociophysics \cite{Galam1997,Bouchaud2013}, specifically to illustrate qualitative campaign effects \cite{Tiwari2021}, but have never been calibrated to an Ising-type model with data on campaign spending. The model employs concepts such as temperature and external fields, which should be understood as effective parameters that summarize complex social, cognitive, and economic processes and not as literally physical quantities.

The aim of this paper is to understand how the interplay between homophily and campaign-following leads to the emergence of \emph{campaign polarization} in ways that can be calibrated to data. Campaign polarization is defined as the normalized difference between the average opinions of voters exposed to each campaign. Low polarization indicates that both groups vote similarly, with decisions mainly shaped by homophily, whereas high polarization means that groups align with campaigns and are less influenced by neighbors. We compute the phase diagram as a function of the ``temperature,'' representing susceptibility to opinion change, and the campaign spending of the two parties.

We focus on the effects of increasing campaign intensity. At low spending, opinions are shaped mainly by homophily, but as campaign influence grows, voter preferences are increasingly influenced by campaign messaging. Key questions are: when does campaign alignment outweigh homophilic similarity, how does polarization evolve at this point, and how does this transition affect social tension and election outcomes? A strength of our model is that it can be calibrated and tested on empirical data; we use the US House elections between 1980 and 2020, particularly. Calibration allows us to infer the ``temperature'' and the critical spending threshold above which campaign polarization rises sharply, and to track how many races exceeded this threshold over four decades. To our knowledge, this is the first time that thermodynamic parameters and a critical spending threshold are directly extracted from historical election data and used to predict levels of polarization in society.

\emph{Election model as an RFIM with a bimodal field.} — We consider $N$ voters with binary opinions $s_i \in {\pm 1}$, representing preference for one of two parties in a bipartisan election. Voters form a social network encoded in the adjacency matrix $A_{ij}$ and interact through homophily, tending to align with neighbors. Each voter also follows one of the two campaigns, modeled by an external field $h_i$ drawn from a bimodal distribution,
\begin{equation}
p(h_i) = p \, \delta(h_i - h^+) + (1-p)\, \delta(h_i + h^{-}) ,
\end{equation}
where $\delta(x)$ is the Dirac delta function and $p \equiv p(h^+)$ is the probability of following the first party’s campaign. With $h^+, h^- \geq 0$, the field takes values $h^+$ or $-h^-$. The system Hamiltonian is
\begin{equation}
H(s_1,\dots,s_N) = -J \sum_{i < j} A_{ij} s_i s_j - \sum_i h_i s_i .
\end{equation}

To solve the model, we apply two approximations: the \emph{configuration model} and a \emph{mean-field approximation} (see Supplemental Material).
Denoting the average magnetization by $m = \langle s_i \rangle$, we arrive at the mean-field Hamiltonian $H^{MF}(s_1,\dots,s_N) = -  \sum_i (\tilde{J} m  + h_i) s_i$, 
where $\tilde{J} = J \langle k \rangle$ and $\langle k \rangle$ is the average node degree. The equilibrium distribution is therefore $p(s|h^\pm) = \exp\left[-\beta (\tilde{J} m \pm h^\pm)s\right]/Z^{\pm}$, where $\beta = (k T)^{-1}$ is the inverse temperature (for the rest of the paper, we set $k=1$), and $Z^{\pm} = 2 \cosh\left[\beta (\tilde{J} m \pm h^{\pm})\right]$ is the partition function. Here, the temperature $T$ represents social volatility—the willingness of individuals to adopt new opinions, even if this increases social stress.

The average magnetization under field $\pm h^{\pm}$ is $m^{\pm} \equiv \langle s \rangle^{\pm}  = \tanh[\beta(\tilde{J} m \pm h^{\pm})]$. The population magnetization is $m = p m^+ + (1-p)m^-$, representing the election outcome ($m=\pm 1$ corresponds to an unanimous result; $m=0$ represents 50:50 split). It satisfies the self-consistency equation,
\begin{equation}\label{eq:m0}
m = p \tanh[\beta(\tilde{J} m + h^{+})] +(1-p)\tanh[\beta(\tilde{J} m - h^-)]\, .
\end{equation}
Equation \eqref{eq:m0} can be alternatively derived from a master-equation of stochastic opinion switching with preference and adaptation terms; the explicit derivation and parameter mapping are given in the Supplemental Material. Setting $m=0$ yields
\begin{equation}\label{eq:zerom}
p \tanh(\beta h^+) = (1-p)\tanh(\beta h^-)\, ,
\end{equation}
which reduces to $h^+=h^-$ for $p=\tfrac{1}{2}$. We define \emph{campaign polarization} as $\pi = \tfrac{1}{2}(m^+ - m^-)$, the difference between average opinions of voters following opposite campaigns. If both groups share the same opinion, $\pi=0$; if they hold opposite views ($m^+=1$, $m^-=-1$), then $\pi=1$.

\emph{Critical parameters.} ---  We first summarize the known results for the symmetric case $p=\tfrac{1}{2}$ and $h^+=h^-\equiv h$. We set $\tilde{J}=1$. As shown in \cite{rfim2010} and in the Supplemental Material, the model exhibits a continuous crossover for $T>1$. For $T<1$, it undergoes a second-order transition at $h_c = T \, \textrm{arctanh}\left(\sqrt{1-T}\right)$. A first-order transition occurs for lower temperatures, with tricritical point $T_t = \tfrac{2}{3}$ and $h_c= \tfrac{2}{3} \, \textrm{arctanh}\left(\tfrac{1}{\sqrt{3}}\right) \approx 0.439$.

We extend this result to the non-symmetric case using Eq.~\eqref{eq:zerom}, which links $h^+$ and $h^-$, and by expanding the self-consistency equation \eqref{eq:m0} around $m=0$. Unlike the symmetric case, the quadratic term does not vanish, yielding the critical curves
\begin{equation}\label{eq:critfieldp}
h_c^+ = T \, \textrm{arctanh}\left(\sqrt{(1-T) \frac{1-p}{p}}\right) \, , 
\end{equation}
\begin{equation}\label{eq:critfieldm}
h_c^- = T \, \textrm{arctanh}\left(\sqrt{(1-T) \frac{p}{1-p}}\right) \, . 
\end{equation}
The full derivation is given in the Supplemental Material.

\begin{figure*}[t]
   \centering
    \includegraphics[width=0.245\linewidth]{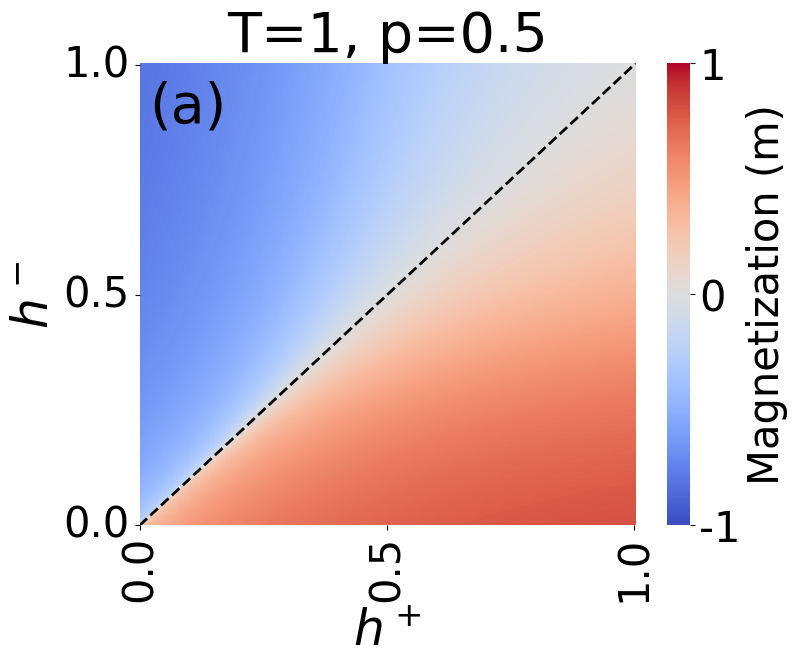}
    \includegraphics[width=0.245\linewidth]{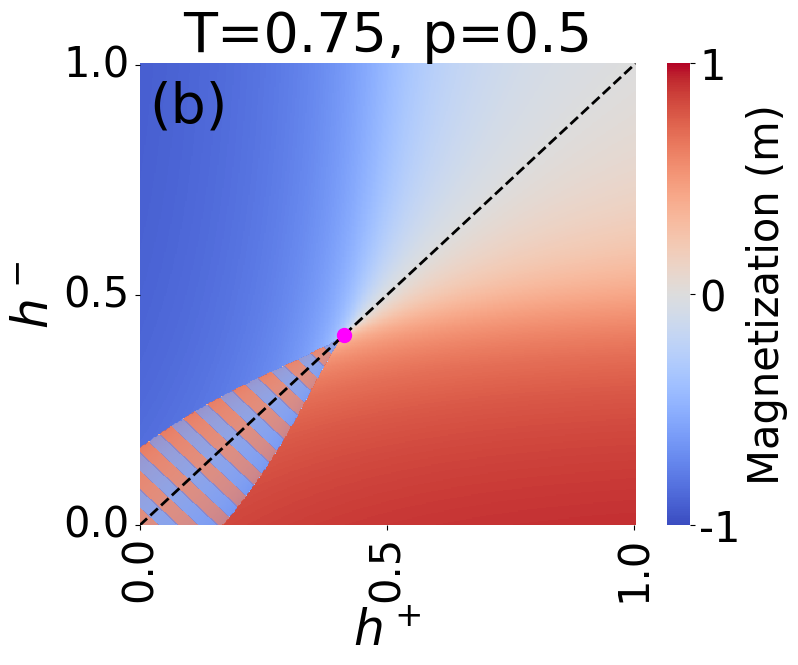}
    \includegraphics[width=0.245\linewidth]{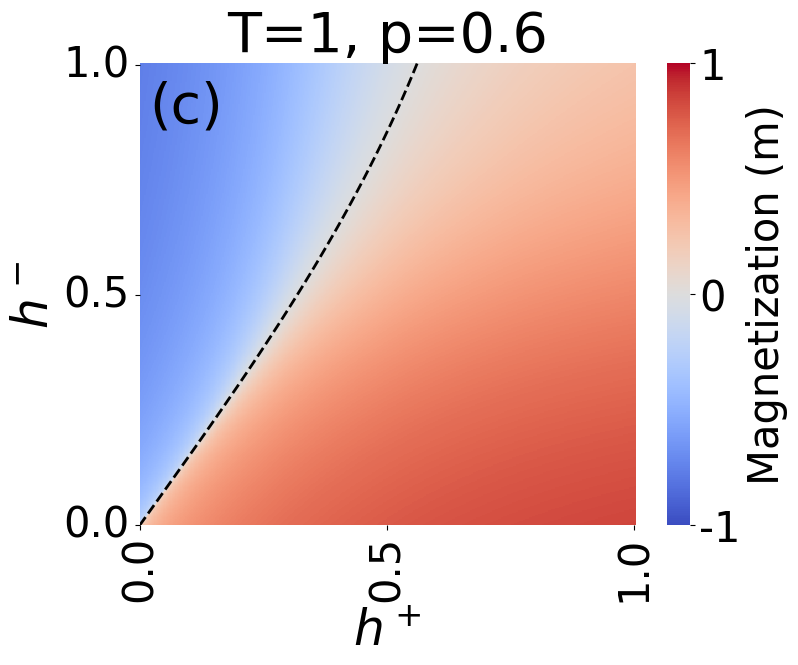}
    \includegraphics[width=0.245\linewidth]{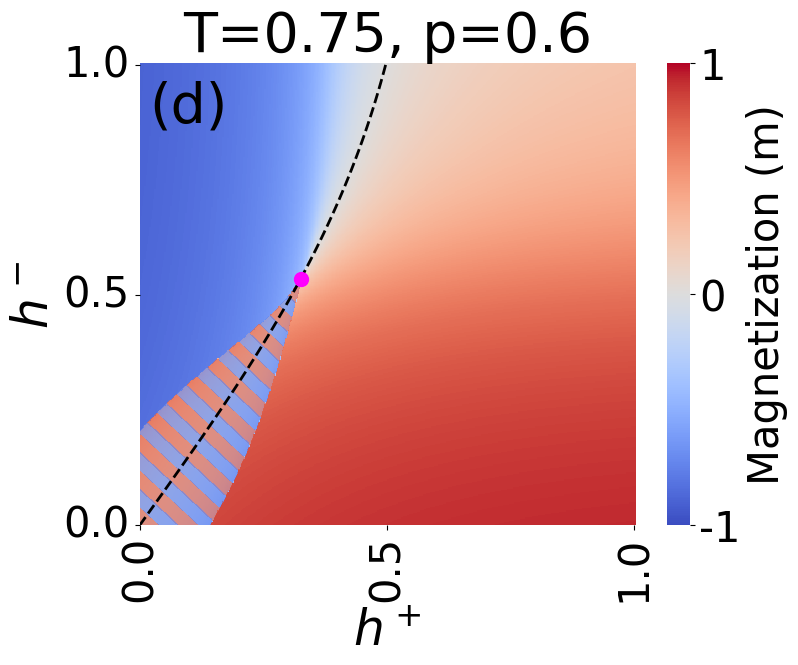}\\
     \includegraphics[width=0.245\linewidth]{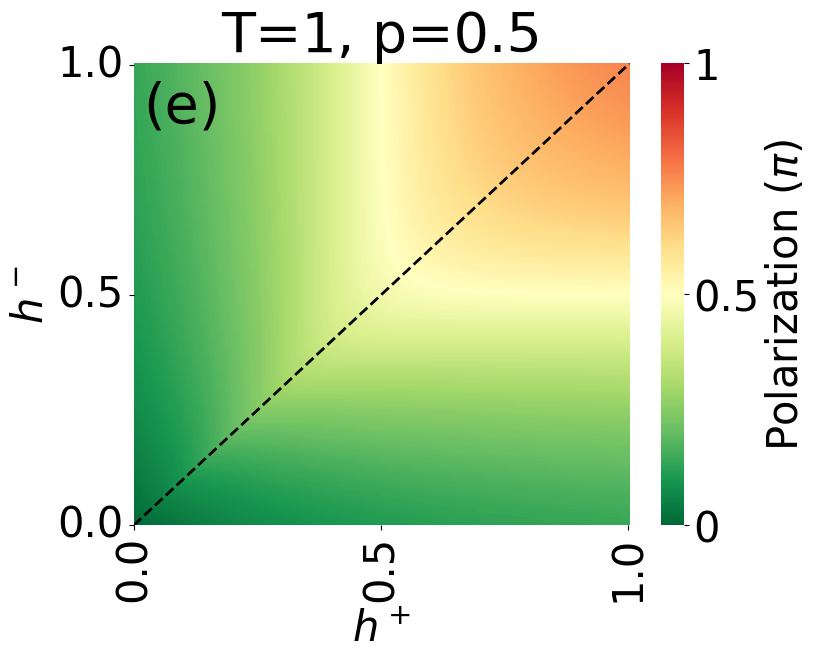}
     \includegraphics[width=0.245\linewidth]{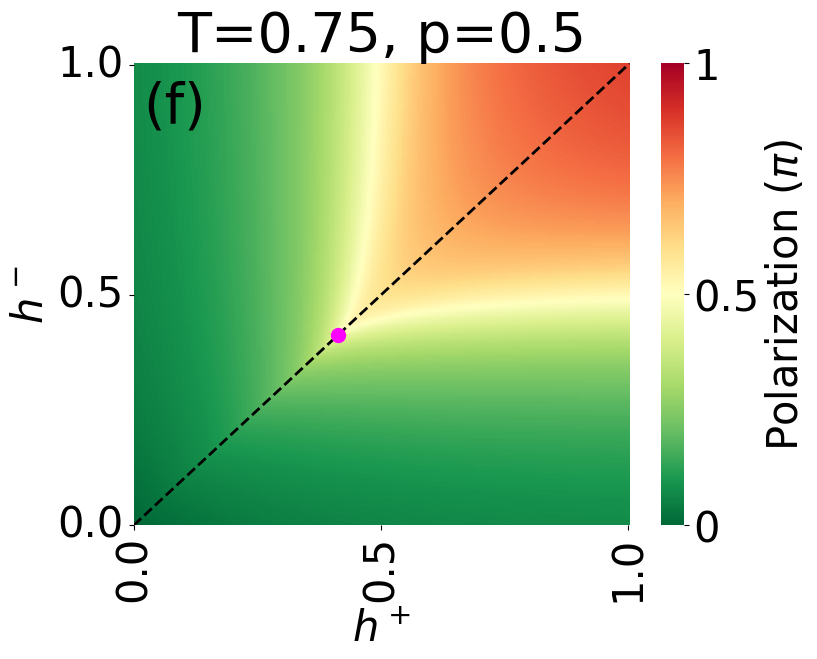}
     \includegraphics[width=0.245\linewidth]{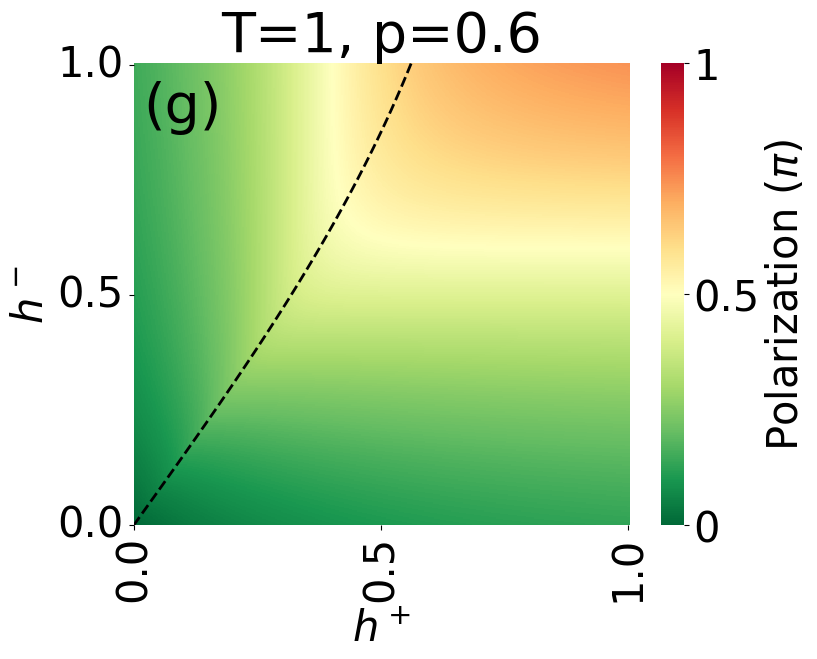}
     \includegraphics[width=0.245\linewidth]{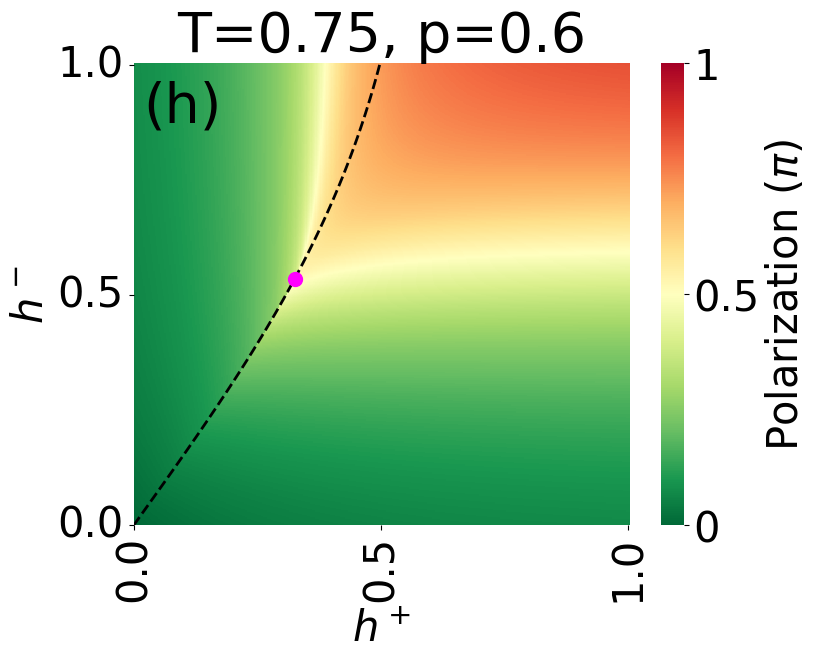}
    \caption{\textbf{Phase diagrams of the election model.} Phase diagram for magnetization, $m$ (a-d), and polarization, $\pi$, (e-h) in the $(h^+,h^-)$ plane for temperature $T=1$ (a,c,e,g), and for $T=0.75$ (b,d,f,h), and a prior probability, $p=0.5$ (a,b,e,f) and $p=0.6$ (c,d,g,h). The black dashed line shows $m=0$. The purple point marks the maximal point of the hysteresis, as derived in the main text. For $T=1$, the expected behavior is that the magnetization is directly affected by the relative strength of the two fields, affected by $p$. For $T=0.75$, we observe a more interesting behavior of the phase diagram. For the case of low field strength, we observe hysteresis (striped region). In both cases, the campaign polarization, $\pi$, increases rapidly when both field strengths exceed a critical value, $h_c$ (red).}
    \label{fig:2} 
\end{figure*}

\emph{Phase diagram.} --- In the $(h^+,h^-)$ plane, we solve Eq.~\eqref{eq:m0} numerically to obtain the phase diagrams shown in Fig.~\ref{fig:2}. For $T \geq 1$, the system has a single stable solution: the candidate with higher campaign spending wins, with the boundary given by Eq.~\eqref{eq:zerom}. For $T<1$ and low fields, a hysteresis region appears around the curve given by Eq.~\eqref{eq:zerom} until reaching the critical values of Eqs.~\eqref{eq:critfieldp} and \eqref{eq:critfieldm}. This implies voter behavior depends on prior states, which we interpret as an \emph{incumbency effect} where officeholders retain an advantage even with lower spending than predicted by Eq.~\eqref{eq:zerom}. The incumbency effect is well documented \cite{Anderson1992,Fowler2018} and considered central to campaign strategy \cite{Druckman2020}.

Campaign polarization $\pi$ remains near zero when at least one external field is weak and increases only when both exceed their critical values, $h^+ \gtrsim h^+_c$ and $h^- \gtrsim h^-_c$. Like magnetization, polarization undergoes a phase transition for $T<1$. In the high-polarization regime, the overall magnetization is nearly constant, $m \approx 2p-1$, as predicted by Eq.~\eqref{eq:zerom} for $h^+,h^- \gg 1$, an effect most pronounced at low $T$. Thus, campaign influence dominates homophily: voters aligned with a campaign tend to vote uniformly for that party, regardless of neighbors. This matches recent results \cite{Hirsch2023} linking strong campaign polarization to \emph{voter extremism}, where voters adopt increasingly extreme positions under intense pressure. Finally, polarization here corresponds to \emph{affective polarization} \cite{iyengar2012affect,lelkes2016mass,iyengar2019origins}, where individuals bond more with their political group than with ideology. Our model captures this as the interplay of ideological and partisan homophily, amplified by campaign intensity.

\begin{figure*}[t]
  \centering
  \begin{minipage}[t]{0.19\textwidth}   \includegraphics[width=\linewidth]{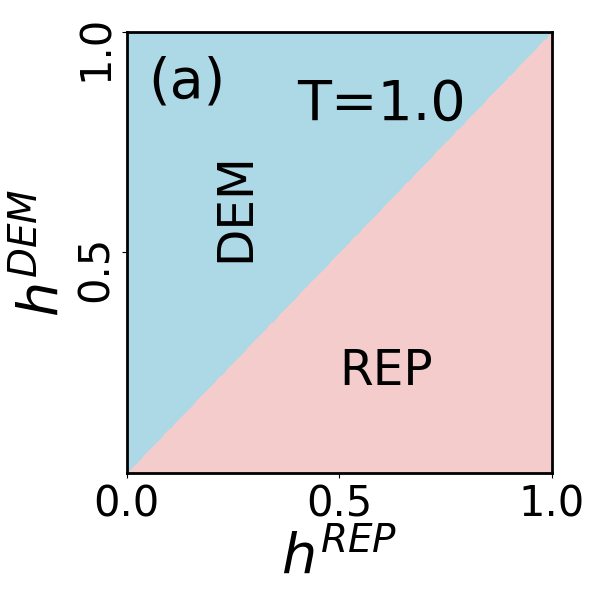}  
 \includegraphics[width=\linewidth]{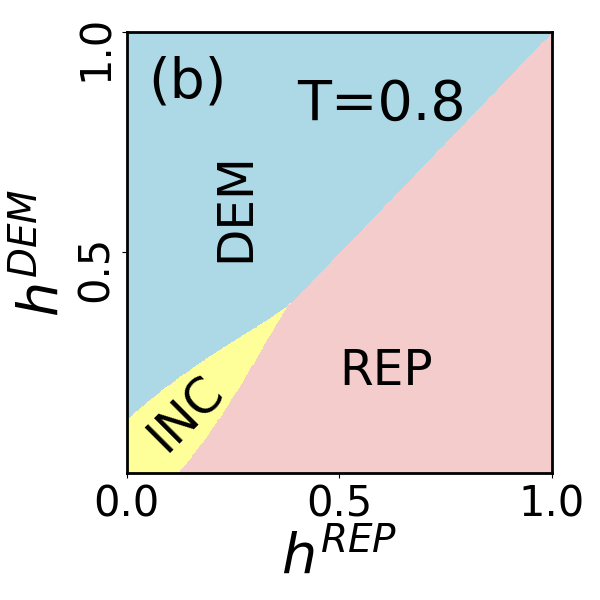}
  \end{minipage}
  \begin{minipage}[c]{0.38\textwidth}  \includegraphics[width=\linewidth, height=\textheight, keepaspectratio]{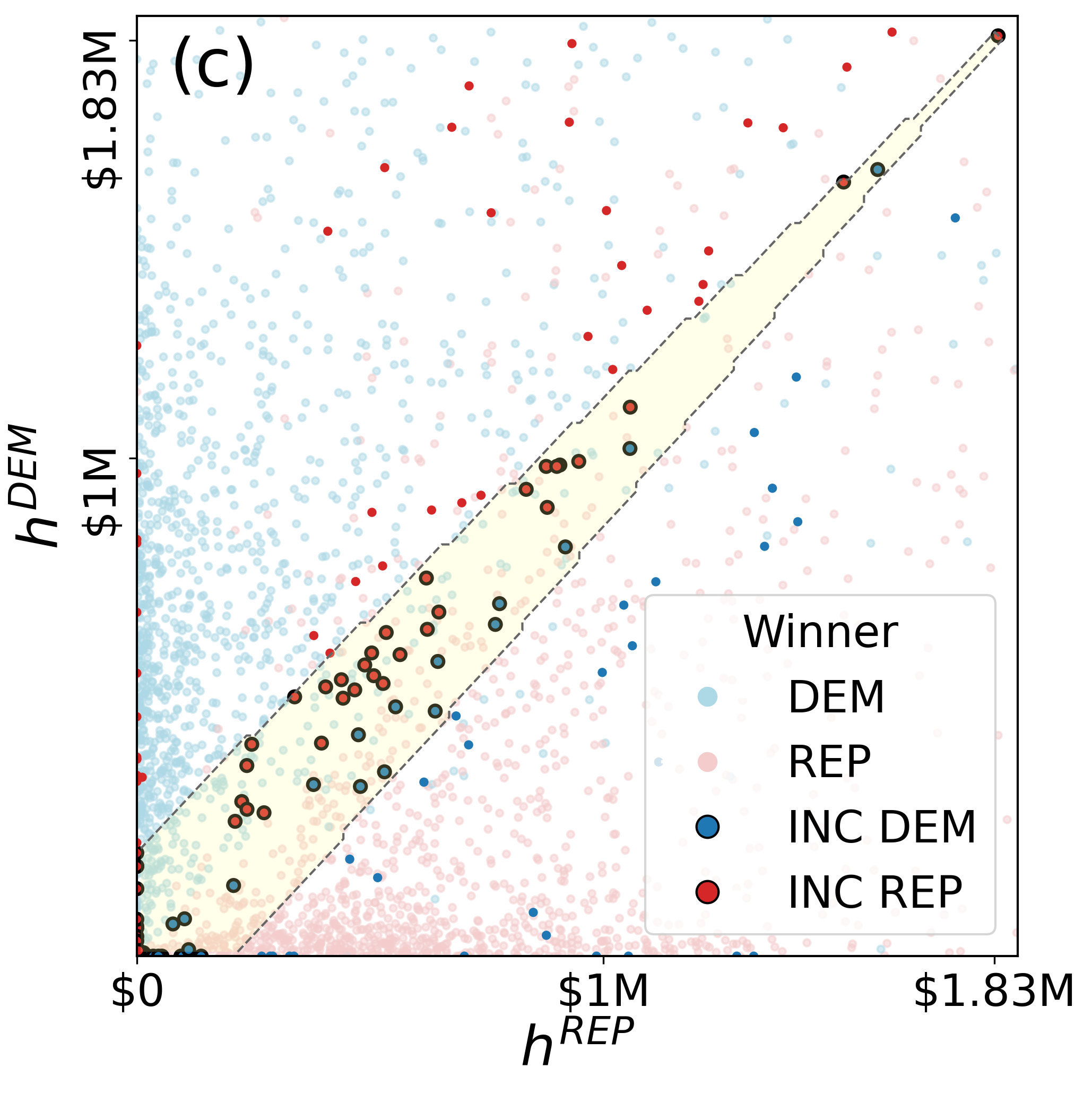}
  \end{minipage}
  \begin{minipage}[t ]{0.38\textwidth}
\includegraphics[width=\linewidth]{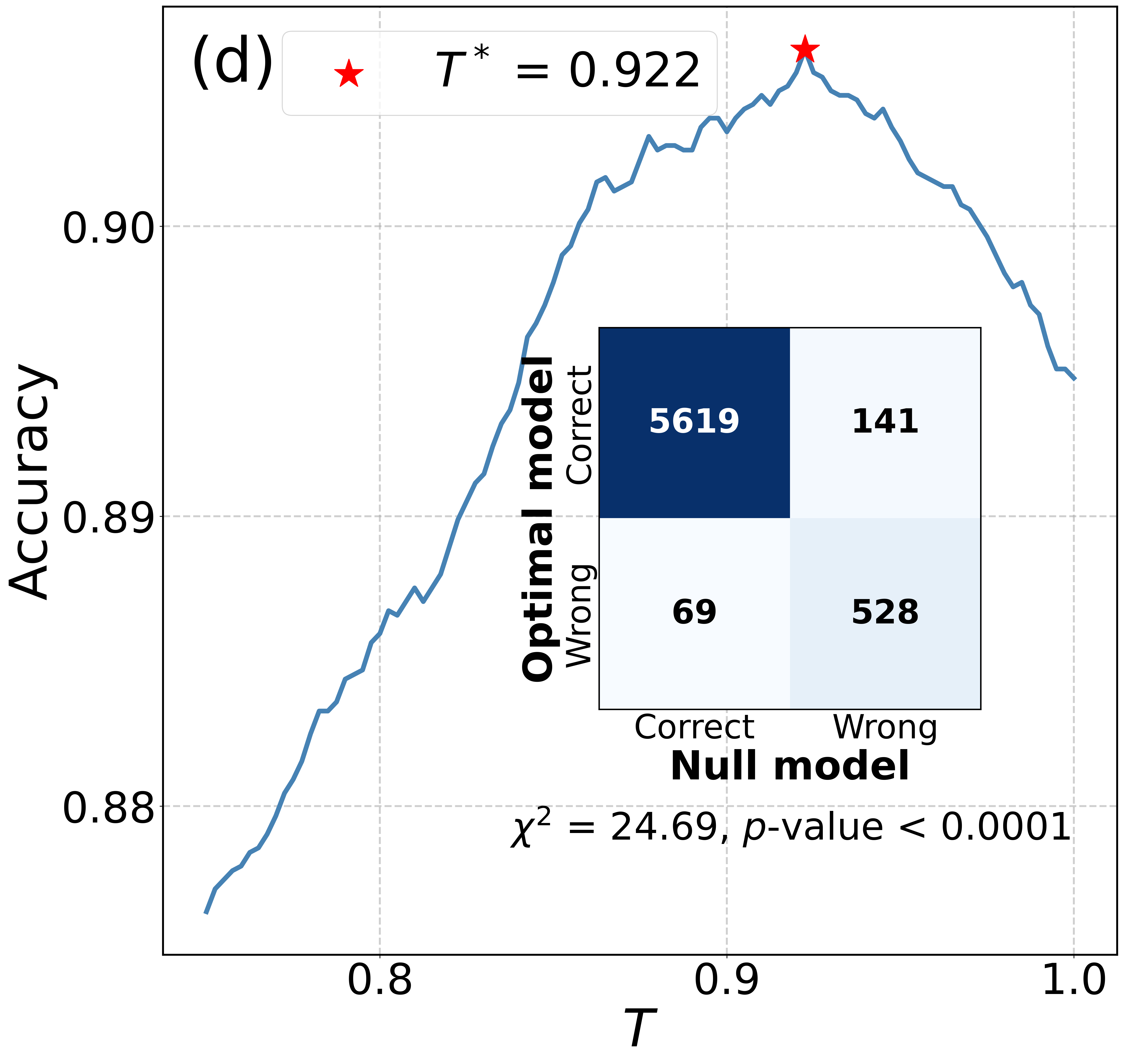}

   \end{minipage}

\caption{\textbf{Estimation of model parameters for US House of Representatives.} 
We compare campaign spending and election results for 6357 of 9135 races between 1980–2020, focusing on close races ($p=0.5 \pm 0.05$). (a) For $T \geq 1$, the classification model (see Supplemental Material) predicts the higher-spending candidate wins. (b) For $T<1$, it predicts an incumbency region (yellow) where incumbents win despite lower spending. (c) Optimal parameters $T$ and $h_c$ are estimated by maximizing classification accuracy across all 6357 races, yielding $T^*=0.922$ and $h_c=\$1.83$M. Cases where incumbents win with lower spending are highlighted, with additional black borders for points in the hysteresis region. The spending diagram is truncated to show the incumbency region (yellow). (d) Accuracy across $T$ is shown, with the maximum marked by a red star. The inset shows a McNemar contingency table comparing the optimal model ($ T=T^*$) to the null model ($T=1$). The McNemar test gives $p<0.0001$, indicating significantly better performance of the optimal model.}
  \label{fig:3}
\end{figure*}

\emph{Calibration to US House election data.} --- To test the model against real data, we analyze all US House campaigns in 435 districts across 21 elections (presidential and midterms) from 1980–2020. Campaign spending and results are publicly available via the Federal Election Commission (FEC) \cite{FEC.gov} and in machine-readable form at \cite{bonica2024dime}. We focus on House races because congressional districts have relatively uniform populations, enabling meaningful spending comparisons. To ensure bipartisan competition, we restrict to races contested solely between Democratic and Republican candidates, excluding those with significant third-party or independent contenders. This yields 6357 races from 9135 in the period. All campaign expenditures are inflation-adjusted to 2020 USD using the Consumer Price Index (CPI). For each race, the campaign share parameter $p$ is set by the previous election result in the same district. For the first election in our dataset (1980), $p$ is taken from the 1978 publicly available results.

To estimate model parameters, we build a classification framework based on the proposed dynamics to predict race winners. Outside the hysteresis region, the winner is determined by the sign of total magnetization; within the hysteresis region, outcomes depend on incumbency when an incumbent is present. This scheme is illustrated in Fig.~\ref{fig:3}(a,b) and detailed in the Supplemental Material.

We first consider the symmetric case $p \approx 0.5$. For $p=0.5$, the classification model predicts the same outcome for all $T \geq 1$: the higher-spending candidate wins (Fig.~\ref{fig:3}(a)). For $T<1$, a hysteresis region emerges (yellow) where incumbency dominates. This region centers near $h^{DEM}=h^{REP}$, with its shape depending on $T$ and $h_c$ (Fig.~\ref{fig:3}(b)). By fitting the empirical boundaries of this region, we infer $T$ and $h_c$. Optimal parameters are estimated by maximizing classification \emph{accuracy} over $T$ and $h_c$, yielding $T^*=0.922$ and $h_c=\$1.83M$ (Fig.~\ref{fig:3}(c)). The figure highlights all closely contested races ($p=0.5 \pm 0.05$), marking cases where incumbents win despite lower spending (black-bordered dots).

\begin{figure*}[h!]
  \centering
  \begin{minipage}[t]{0.19\textwidth}   \includegraphics[width=\linewidth]{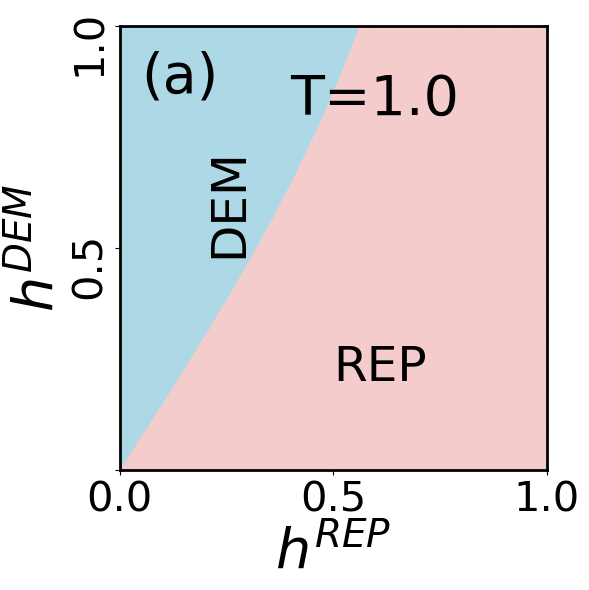}  
 \includegraphics[width=\linewidth]{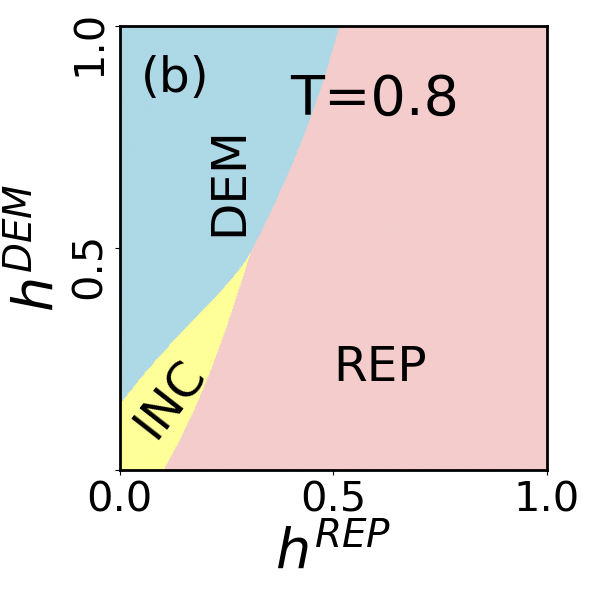}
  \end{minipage}
  \begin{minipage}[c]{0.38\textwidth}  \includegraphics[width=\linewidth, height=\textheight, keepaspectratio]{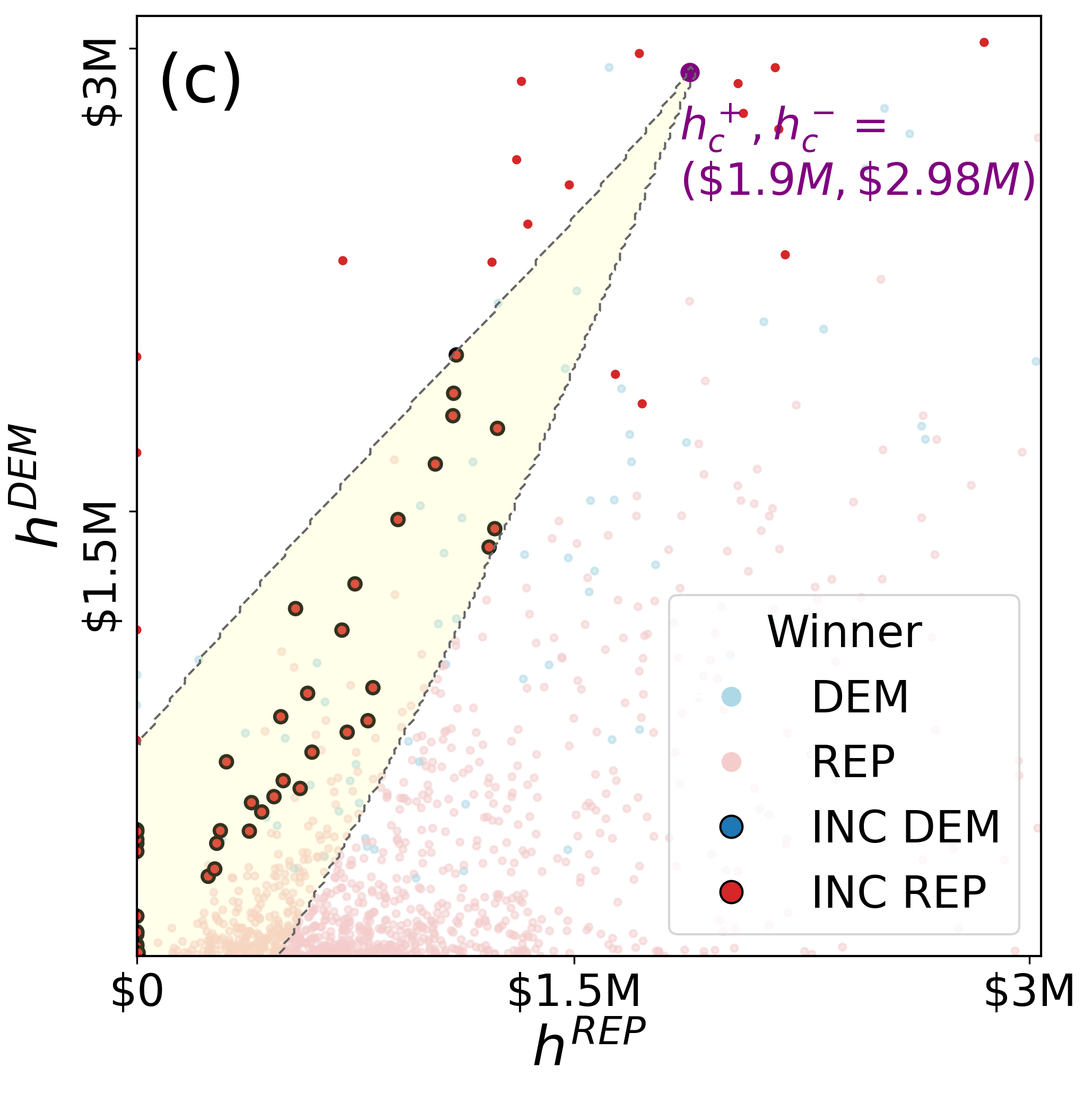}
  \end{minipage}
  \begin{minipage}[t ]{0.38\textwidth}
\includegraphics[width=\linewidth]{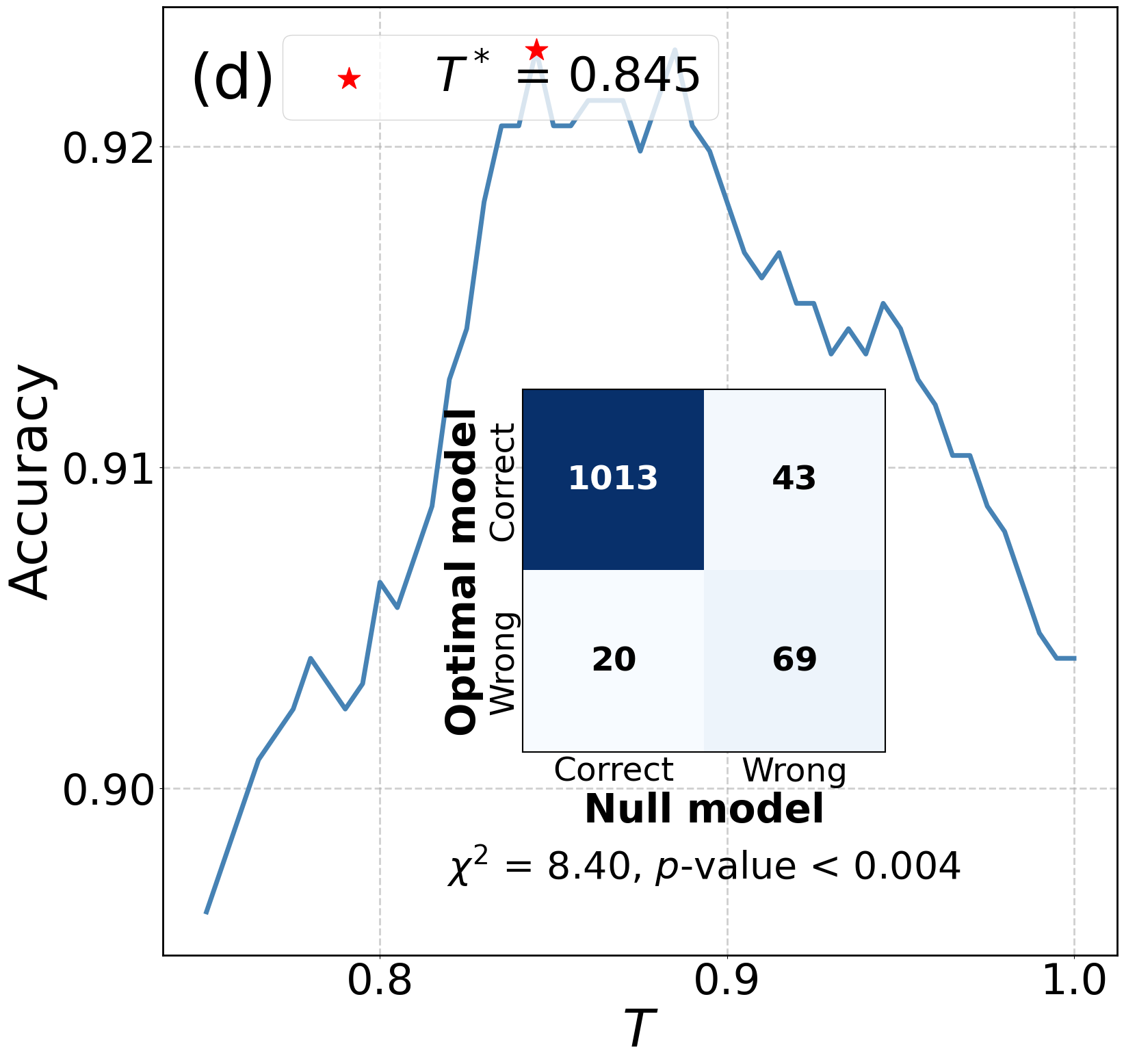}

   \end{minipage}
    \caption{{\textbf{Estimation of model parameters for US House of Representatives for Republican-leaning races ($p=0.6 \pm 0.05$).} Similarly to \ref{fig:3} in the main text, we use the classification model for $p=0.6$ to estimate the parameters of the model. (a) Classification model for $T=1$ without hysteresis. (b) Classification model for $T<1$ with incumbent region. (c) The plot of the election results with the campaign spending, with the incumbent region predicted by the optimal model. (d) The accuracy of the classification model in a range of temperatures; the star denotes the model with the best accuracy, corresponding to $T^\star=0.845$.}}
    \label{fig:rep}
\end{figure*}

\begin{figure*}
 \centering
  \begin{minipage}[t]{0.19\textwidth}   \includegraphics[width=\linewidth]{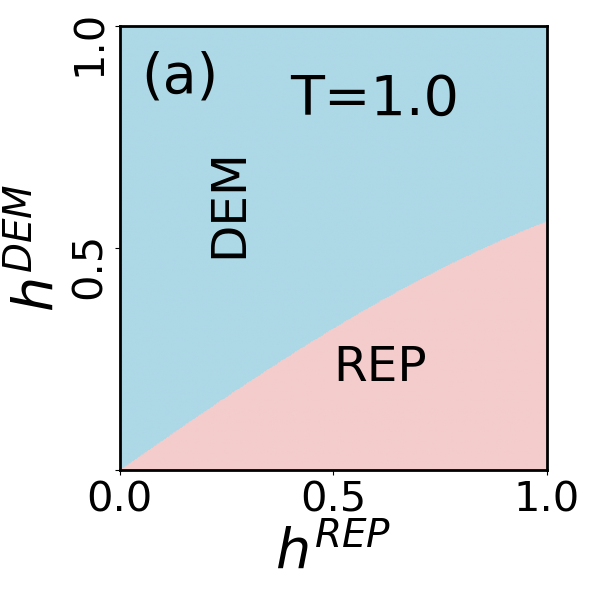}  
  
\includegraphics[width=\linewidth]{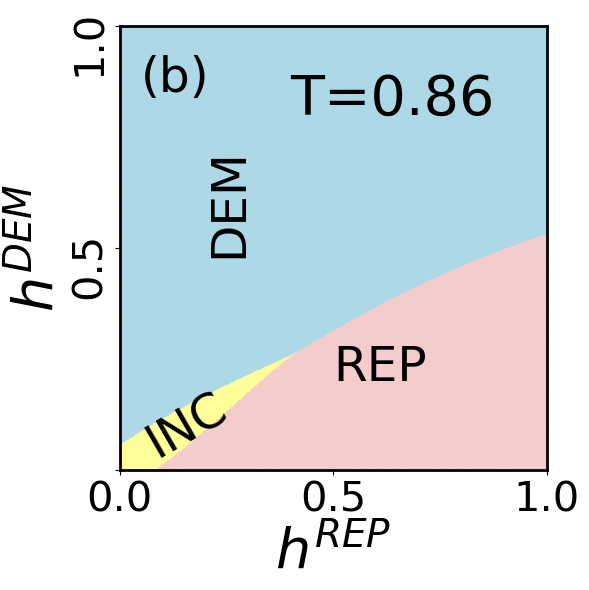}
  \end{minipage}
  \begin{minipage}[c]{0.38\textwidth}  \includegraphics[width=\linewidth, height=\textheight, keepaspectratio]{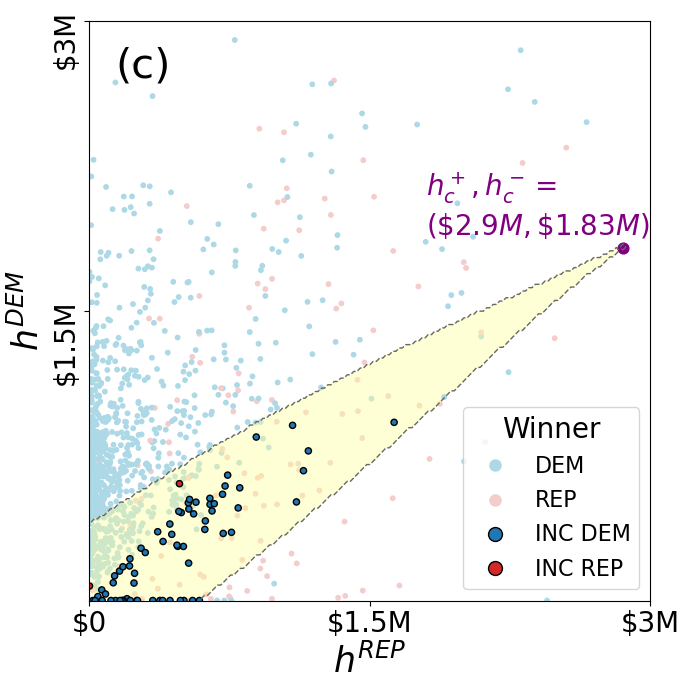}
  \end{minipage}
  \begin{minipage}[t ]{0.38\textwidth}
\includegraphics[width=\linewidth]{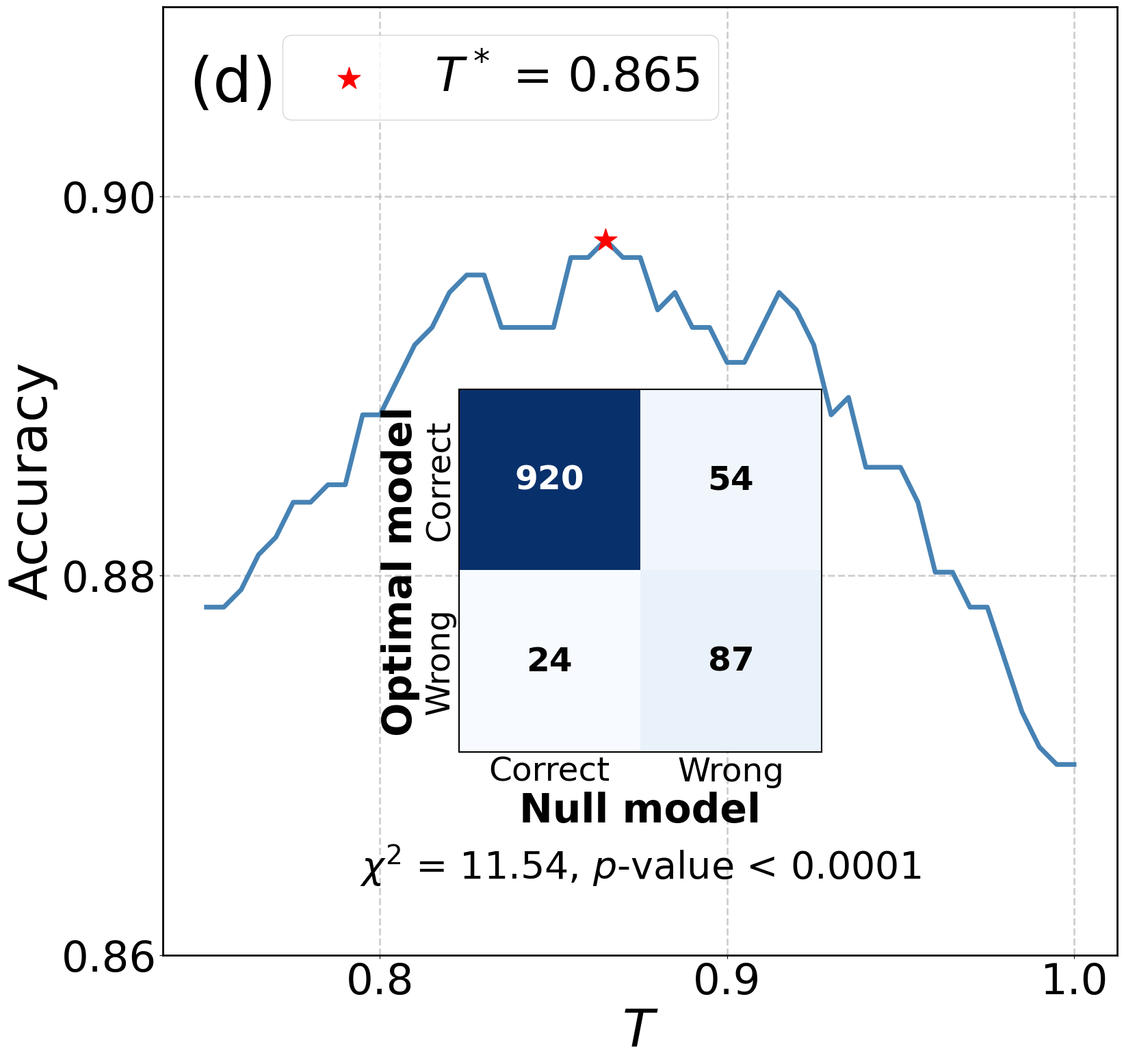}

   \end{minipage}
    \caption{{\textbf{Estimation of model parameters for US House of Representatives for Democrat-leaning races ($p=0.4 \pm 0.05$).} We use the classification model for $p=0.4$ to estimate the parameters of the model. (a) Classification model for $T=1$ without hysteresis. (b) Classification model for $T<1$ with incumbent region. (c) The plot of the election results with the campaign spending, with the incumbent region predicted by the optimal model. (d) The accuracy of the classification model in a range of temperatures; the star denotes the model with the best accuracy, corresponding to $T^\star=0.865$.}}
    \label{fig:dem}
\end{figure*}

In the End Matter, we present a similar analysis for races with $p \approx 0.6$, i.e., Republican-leaning races ($p=0.6 \pm 0.05$, Fig. \ref{fig:rep}) and $p \approx 0.4$, i.e., Democrat-leaning races ($p= 0.4 \pm 0.05$, Fig. \ref{fig:dem}). Applied to these two subsets, the parameters shift slightly ($T^*=0.845$ for Republican-leaning races, $T^*=0.865$ for Democrat-leaning races, $h_c \sim \$2M$), but the qualitative behavior --- hysteresis and emergence of polarization --- remains unchanged.

Figure~\ref{fig:3}(d) shows classification accuracy as a function of temperature, with $h_c$ chosen at each $T$ to maximize accuracy. To test statistical significance against a baseline, we compare the optimal classifier to a null model without hysteresis ($T \geq 1$) using the McNemar test (see Supplemental Material). The inset shows the contingency table of correctly and incorrectly classified results for both models.

We find a statistically significant improvement in accuracy ($\chi^2=24.69$, p-value $p=6.76\cdot10^{-7}$), though the absolute gain is small. This reflects that in most races incumbents outspend challengers, so both the null model ($T \geq 1$) and optimal model ($T=T^\star$) yield similar predictions. This aligns with prior findings that incumbency strengthens fundraising, with incumbents typically attracting more support \cite{Fouirnaies2014}.

\begin{figure*}[h]
  \centering
  \begin{minipage}[c]{0.38\textwidth}
    \includegraphics[width=\linewidth]{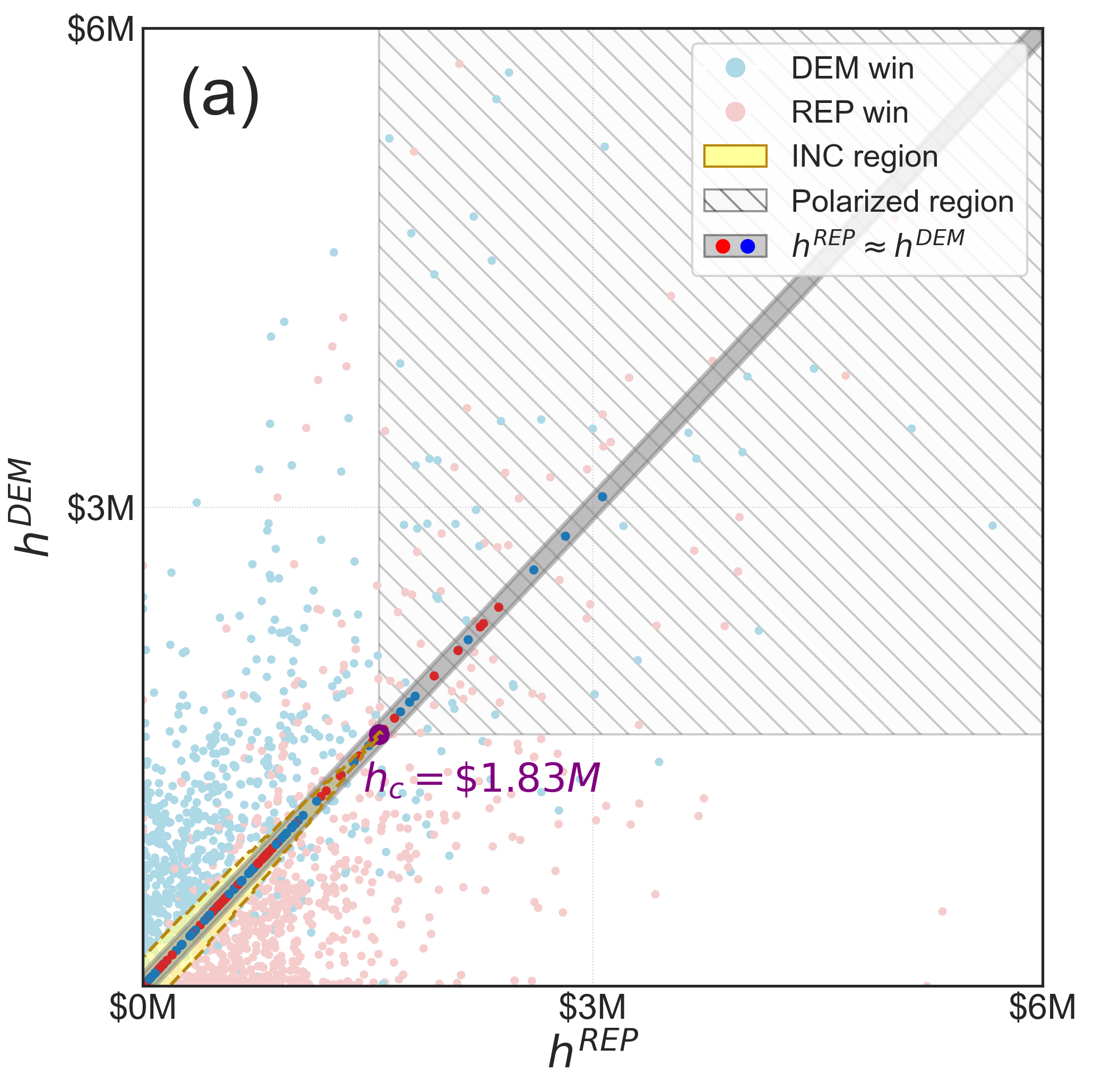}
  \end{minipage}
  \begin{minipage}[t]{0.38\textwidth}
    \includegraphics[width=\linewidth, height=0.48\textheight, keepaspectratio]{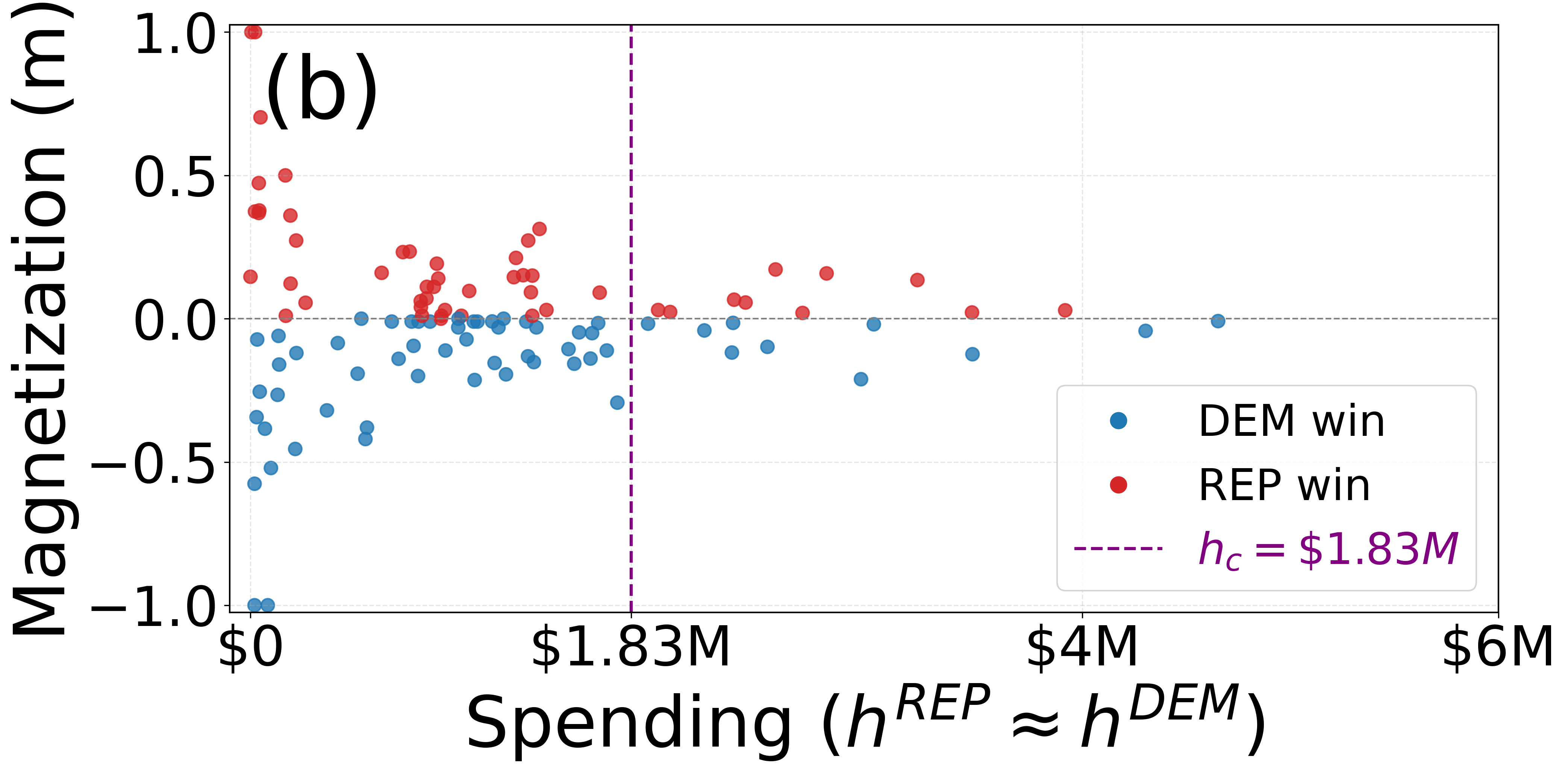}

   \includegraphics[width=0.97\linewidth, height=\textheight, keepaspectratio]{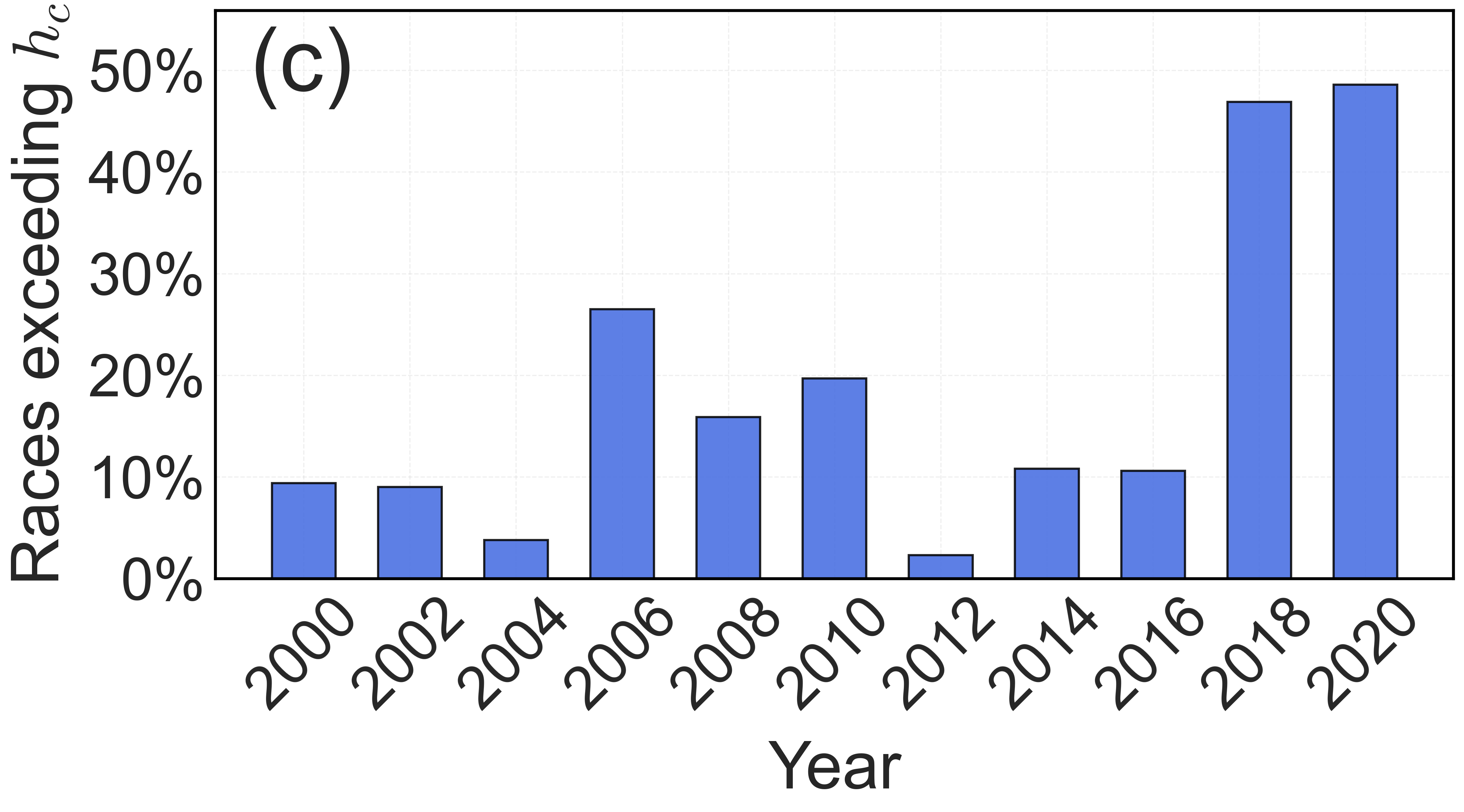}
  \end{minipage}
\caption{\textbf{Emergence of campaign polarization in the US House of Representatives elections.}
We compare campaign spending and election results for races from 1980–2020, focusing on close contests with $p = 0.5 \pm 0.05$.
(a) Phase diagram as in Fig.~\ref{fig:3}, now covering the full range of spending, including the polarized region; the gray area marks races with close spending.
(b) Election outcomes near $h^{DEM} \approx h^{REP}$, where $|h^{DEM}-h^{REP}|<\$100,000$. The $x$-axis shows average spending $\tfrac{1}{2}(h^{DEM}+h^{REP})$. Below $h_c$, outcomes are mostly decisive, with only 39\% close races ($|m|<0.1$) and stronger incumbency effects. Above $h_c$, over 70\% of outcomes are near 50:50, consistent with RFIM predictions.
(c) Percentage of close races where both campaigns exceed $h_c$, i.e., within the polarized region $\pi$. This share rises sharply in 2018 and 2020.
}
  \label{fig:4}
\end{figure*}

In the End Matter, we further extend our analysis to the emergence of polarization. By estimating the critical spending threshold $h_c$, we test how many races fall in the polarization region, i.e., when both parties exceed $h_c$ (Fig.~\ref{fig:4}). Panel (a) shows the full spending region including the polarized area, while panel (b) presents results for $p \approx 0.5$. At low spending, outcomes are decisive; above $h_c$, results cluster near 50:50, consistent with the prediction that for $p \approx 0.5$ most outcomes in the polarized region satisfy $m \approx 0$. Panel (c) compares the number of races where both campaigns exceeded $h_c$ across election cycles, revealing a sharp rise in 2018 and 2020. While these results rest on model predictions and need further empirical validation, the trend matches recent observations of rising campaign polarization \cite{pewresearch_report}.

To further test the robustness of our approach, we present two additional analyses in the Supplemental Material: in the first analysis, we divide the dataset into four decades and estimate the parameters for each decade separately. We observe that while the temperature slightly decreases over time, the critical spending slightly increases. Second, we compare our approach to a machine learning approach based on a support vector machine classification model. We show that the SVM can estimate the incumbency region (with slightly lower test accuracy); however, it cannot provide additional interpretation of the model, such as the presence of a polarization region.

\emph{Discussion.} --- We introduced a simple election model combining two mechanisms influencing voter decisions: homophily, i.e., interactions with family, friends, and close contacts, and political campaign influence. Despite its simplicity, the model can be calibrated with real-world data (US House elections) and reproduces rich behavior absent in earlier work. The fact that voters typically follow only one campaign leads to a phase transition: campaign polarization rises sharply once both parties exceed a critical spending threshold. In this regime, most outcomes are 50:50 when $p \approx 0.5$ (swing states), regardless of the detailed spending. For biased states ($p \not\approx 0.5$), the model allows us to estimate the challenger’s minimum spending above which 50:50 outcomes become possible.

We identified a hysteresis region in the phase diagram that leads to an incumbency effect, enabling quantitative assessment of incumbents’ advantage. The model shows that challengers must overcome an initial threshold of about $\$140,000$ even if incumbents spend nothing. This barrier decreases with higher overall spending but remains significant: for instance, when the incumbent spends $0.5 h_c$ ($\approx \$900,000$), the challenger still faces a disadvantage of about 20\% of total campaign cost. This quantifies a structural incumbency advantage beyond candidate-specific factors. By quantifying incumbency strength, we estimate the effective temperature of the social system and identify a polarization threshold. Notably, even without a universal definition of ``temperature'' in social contexts, it can be inferred indirectly from observable critical phenomena such as hysteresis and field-driven phase transitions.

Several studies reported a decline in incumbency strength in recent decades \cite{Jacobson2015}, raising the question of whether this relates to shifts in effective ``temperature''. It remains open whether this temperature is universal or varies across elections, contexts, or regions. Since incumbency is central to electoral strategy \cite{Stone2010}, further work should examine how decisions such as planned retirements shape challengers’ prospects by mitigating incumbents’ inherent advantage.

The model can be naturally extended in several ways. Multipartisan systems can be described by an extension to a double-random field Potts model \cite{Blankschtein1984} with more than two opinion states. More realistic scenarios could be added through extensions to heterogeneous friendship networks, explicit party membership, or mechanisms such as primaries. Homophily may also vary across ties—for instance, individuals might ignore co-workers’ views but adopt those of parents or close friends. Such modifications would better capture electoral complexity. Moreover, campaign intensity may not scale linearly with spending, as noted in earlier work \cite{Jacobson1978,Jacobson1990}, suggesting future models should consider nonlinear or context-dependent effects.

Despite its simplicity, the model predicts complex game-theoretic behavior. Strategic aspects of campaigns have been studied from this perspective \cite{Erikson2000,Schnakenberg2021}, and our results deepen this understanding. The model suggests that candidates may rationally raise spending to induce polarization; once reached, it is difficult for opponents to reverse, locking in an advantage. This is especially effective in partisan districts, where dominant candidates benefit from driving the electorate into a polarized phase. Yet this strategy carries social costs: polarization erodes ties across divides and exacerbates fragmentation.
A key implication of the model is that regulatory interventions, such as caps on campaign spending, may be needed to prevent undesirable outcomes. While spending limits are under renewed discussion \cite{Avis2022}, our results add a novel perspective by emphasizing the broader societal costs of unbounded campaign intensity.

Note that similar results were found in the context of polarization in the US Congress itself \cite{luEvolutionPolarizationLegislative2019}. The study shows that the polarization in the US Congress increased after the 2010 Supreme Court approval of Super PACs, which enabled an increase in donor influence. Also, other aspects such as social connectivity can increase polarization, not only in the context of elections \cite{Thurner2025,Pham2025}.

Elections are not the only context in which homophily competes with external influence. Similar dynamics appear in marketing, where peer effects and advertising shape consumer behavior. A classic case is the Coca-Cola vs. Pepsi rivalry \cite{Golan2000}, showing how competing campaigns polarize preferences. Our framework illustrates how polarization generally arises from the tension between social ties and persuasive efforts.

\emph{Acknowledgments.} --- We thank Mirta Galesic and Henrik Olsson for helpful discussions. We also thank three anonymous referees for their constructive comments. We acknowledge support from the Austrian Science Fund (FWF) under Grants No. 10.55776/P34994 and EFP5 ReMASS, funding from the Austrian Federal Ministry for Climate Action, Environment, Energy, Mobility, Innovation, and Technology under GZ 2023-0.841.266, through the Postdoc Program for Complexity Science and Data Competence.

\emph{Data availability statement.} --- he data that support the findings of this article are openly available \cite{FEC.gov,bonica2024dime}.

\bibliography{bibliography}

\begin{thebibliography}{74}%
\makeatletter
\providecommand \@ifxundefined [1]{%
 \@ifx{#1\undefined}
}%
\providecommand \@ifnum [1]{%
 \ifnum #1\expandafter \@firstoftwo
 \else \expandafter \@secondoftwo
 \fi
}%
\providecommand \@ifx [1]{%
 \ifx #1\expandafter \@firstoftwo
 \else \expandafter \@secondoftwo
 \fi
}%
\providecommand \natexlab [1]{#1}%
\providecommand \enquote  [1]{``#1''}%
\providecommand \bibnamefont  [1]{#1}%
\providecommand \bibfnamefont [1]{#1}%
\providecommand \citenamefont [1]{#1}%
\providecommand \href@noop [0]{\@secondoftwo}%
\providecommand \href [0]{\begingroup \@sanitize@url \@href}%
\providecommand \@href[1]{\@@startlink{#1}\@@href}%
\providecommand \@@href[1]{\endgroup#1\@@endlink}%
\providecommand \@sanitize@url [0]{\catcode `\\12\catcode `\$12\catcode `\&12\catcode `\#12\catcode `\^12\catcode `\_12\catcode `\%12\relax}%
\providecommand \@@startlink[1]{}%
\providecommand \@@endlink[0]{}%
\providecommand \url  [0]{\begingroup\@sanitize@url \@url }%
\providecommand \@url [1]{\endgroup\@href {#1}{\urlprefix }}%
\providecommand \urlprefix  [0]{URL }%
\providecommand \Eprint [0]{\href }%
\providecommand \doibase [0]{https://doi.org/}%
\providecommand \selectlanguage [0]{\@gobble}%
\providecommand \bibinfo  [0]{\@secondoftwo}%
\providecommand \bibfield  [0]{\@secondoftwo}%
\providecommand \translation [1]{[#1]}%
\providecommand \BibitemOpen [0]{}%
\providecommand \bibitemStop [0]{}%
\providecommand \bibitemNoStop [0]{.\EOS\space}%
\providecommand \EOS [0]{\spacefactor3000\relax}%
\providecommand \BibitemShut  [1]{\csname bibitem#1\endcsname}%
\let\auto@bib@innerbib\@empty
\bibitem [{\citenamefont {Mantegna}\ and\ \citenamefont {Stanley}(1995)}]{Mantegna1995}%
  \BibitemOpen
  \bibfield  {author} {\bibinfo {author} {\bibfnamefont {R.~N.}\ \bibnamefont {Mantegna}}\ and\ \bibinfo {author} {\bibfnamefont {H.~E.}\ \bibnamefont {Stanley}},\ }\bibfield  {title} {\bibinfo {title} {Scaling behaviour in the dynamics of an economic index},\ }\href {https://doi.org/10.1038/376046a0} {\bibfield  {journal} {\bibinfo  {journal} {Nature}\ }\textbf {\bibinfo {volume} {376}},\ \bibinfo {pages} {46} (\bibinfo {year} {1995})}\BibitemShut {NoStop}%
\bibitem [{\citenamefont {Gabaix}\ \emph {et~al.}(2003)\citenamefont {Gabaix}, \citenamefont {Gopikrishnan}, \citenamefont {Plerou},\ and\ \citenamefont {Stanley}}]{Gabaix2003}%
  \BibitemOpen
  \bibfield  {author} {\bibinfo {author} {\bibfnamefont {X.}~\bibnamefont {Gabaix}}, \bibinfo {author} {\bibfnamefont {P.}~\bibnamefont {Gopikrishnan}}, \bibinfo {author} {\bibfnamefont {V.}~\bibnamefont {Plerou}},\ and\ \bibinfo {author} {\bibfnamefont {H.~E.}\ \bibnamefont {Stanley}},\ }\bibfield  {title} {\bibinfo {title} {A theory of power-law distributions in financial market fluctuations},\ }\href {https://doi.org/10.1038/nature01624} {\bibfield  {journal} {\bibinfo  {journal} {Nature}\ }\textbf {\bibinfo {volume} {423}},\ \bibinfo {pages} {267} (\bibinfo {year} {2003})}\BibitemShut {NoStop}%
\bibitem [{\citenamefont {Bouchaud}(2023)}]{Bouchaud2023}%
  \BibitemOpen
  \bibfield  {author} {\bibinfo {author} {\bibfnamefont {J.-P.}\ \bibnamefont {Bouchaud}},\ }\bibfield  {title} {\bibinfo {title} {From statistical physics to social sciences: the pitfalls of multi-disciplinarity},\ }\href {https://doi.org/10.1088/2632-072X/ad104a} {\bibfield  {journal} {\bibinfo  {journal} {J. Phys. Complex.}\ }\textbf {\bibinfo {volume} {4}},\ \bibinfo {pages} {041001} (\bibinfo {year} {2023})}\BibitemShut {NoStop}%
\bibitem [{\citenamefont {Helbing}\ and\ \citenamefont {Moln\'ar}(1995)}]{Helbing1995}%
  \BibitemOpen
  \bibfield  {author} {\bibinfo {author} {\bibfnamefont {D.}~\bibnamefont {Helbing}}\ and\ \bibinfo {author} {\bibfnamefont {P.}~\bibnamefont {Moln\'ar}},\ }\bibfield  {title} {\bibinfo {title} {Social force model for pedestrian dynamics},\ }\href {https://doi.org/10.1103/PhysRevE.51.4282} {\bibfield  {journal} {\bibinfo  {journal} {Phys. Rev. E}\ }\textbf {\bibinfo {volume} {51}},\ \bibinfo {pages} {4282} (\bibinfo {year} {1995})}\BibitemShut {NoStop}%
\bibitem [{\citenamefont {Karamouzas}\ \emph {et~al.}(2014)\citenamefont {Karamouzas}, \citenamefont {Skinner},\ and\ \citenamefont {Guy}}]{Karamouzas2014}%
  \BibitemOpen
  \bibfield  {author} {\bibinfo {author} {\bibfnamefont {I.}~\bibnamefont {Karamouzas}}, \bibinfo {author} {\bibfnamefont {B.}~\bibnamefont {Skinner}},\ and\ \bibinfo {author} {\bibfnamefont {S.~J.}\ \bibnamefont {Guy}},\ }\bibfield  {title} {\bibinfo {title} {Universal power law governing pedestrian interactions},\ }\href {https://doi.org/10.1103/PhysRevLett.113.238701} {\bibfield  {journal} {\bibinfo  {journal} {Phys. Rev. Lett.}\ }\textbf {\bibinfo {volume} {113}},\ \bibinfo {pages} {238701} (\bibinfo {year} {2014})}\BibitemShut {NoStop}%
\bibitem [{\citenamefont {Lansing}\ \emph {et~al.}(2017)\citenamefont {Lansing}, \citenamefont {Thurner}, \citenamefont {Chung}, \citenamefont {Coudurier-Curveur}, \citenamefont {Çağil Karakaş}, \citenamefont {Fesenmyer},\ and\ \citenamefont {Chew}}]{Lansing2017}%
  \BibitemOpen
  \bibfield  {author} {\bibinfo {author} {\bibfnamefont {J.~S.}\ \bibnamefont {Lansing}}, \bibinfo {author} {\bibfnamefont {S.}~\bibnamefont {Thurner}}, \bibinfo {author} {\bibfnamefont {N.~N.}\ \bibnamefont {Chung}}, \bibinfo {author} {\bibfnamefont {A.}~\bibnamefont {Coudurier-Curveur}}, \bibinfo {author} {\bibnamefont {Çağil Karakaş}}, \bibinfo {author} {\bibfnamefont {K.~A.}\ \bibnamefont {Fesenmyer}},\ and\ \bibinfo {author} {\bibfnamefont {L.~Y.}\ \bibnamefont {Chew}},\ }\bibfield  {title} {\bibinfo {title} {Adaptive self-organization of bali’s ancient rice terraces},\ }\href {https://doi.org/10.1073/pnas.1605369114} {\bibfield  {journal} {\bibinfo  {journal} {PNAS}\ }\textbf {\bibinfo {volume} {114}},\ \bibinfo {pages} {6504} (\bibinfo {year} {2017})}\BibitemShut {NoStop}%
\bibitem [{\citenamefont {Gandica}\ \emph {et~al.}(2021)\citenamefont {Gandica}, \citenamefont {Lansing}, \citenamefont {Chung}, \citenamefont {Thurner},\ and\ \citenamefont {Chew}}]{Gandica2021}%
  \BibitemOpen
  \bibfield  {author} {\bibinfo {author} {\bibfnamefont {Y.}~\bibnamefont {Gandica}}, \bibinfo {author} {\bibfnamefont {J.~S.}\ \bibnamefont {Lansing}}, \bibinfo {author} {\bibfnamefont {N.~N.}\ \bibnamefont {Chung}}, \bibinfo {author} {\bibfnamefont {S.}~\bibnamefont {Thurner}},\ and\ \bibinfo {author} {\bibfnamefont {L.~Y.}\ \bibnamefont {Chew}},\ }\bibfield  {title} {\bibinfo {title} {Bali's ancient rice terraces: A hamiltonian approach},\ }\href {https://doi.org/10.1103/PhysRevLett.127.168301} {\bibfield  {journal} {\bibinfo  {journal} {Phys. Rev. Lett.}\ }\textbf {\bibinfo {volume} {127}},\ \bibinfo {pages} {168301} (\bibinfo {year} {2021})}\BibitemShut {NoStop}%
\bibitem [{\citenamefont {Lee}\ \emph {et~al.}(2015)\citenamefont {Lee}, \citenamefont {Broedersz},\ and\ \citenamefont {Bialek}}]{Lee2015}%
  \BibitemOpen
  \bibfield  {author} {\bibinfo {author} {\bibfnamefont {E.~D.}\ \bibnamefont {Lee}}, \bibinfo {author} {\bibfnamefont {C.~P.}\ \bibnamefont {Broedersz}},\ and\ \bibinfo {author} {\bibfnamefont {W.}~\bibnamefont {Bialek}},\ }\bibfield  {title} {\bibinfo {title} {Statistical mechanics of the us supreme court},\ }\href {https://doi.org/10.1007/s10955-015-1253-6} {\bibfield  {journal} {\bibinfo  {journal} {J. Stat. Phys.}\ }\textbf {\bibinfo {volume} {160}},\ \bibinfo {pages} {275} (\bibinfo {year} {2015})}\BibitemShut {NoStop}%
\bibitem [{\citenamefont {Galam}(1999)}]{Galam1999}%
  \BibitemOpen
  \bibfield  {author} {\bibinfo {author} {\bibfnamefont {S.}~\bibnamefont {Galam}},\ }\bibfield  {title} {\bibinfo {title} {Application of statistical physics to politics},\ }\href {https://doi.org/https://doi.org/10.1016/S0378-4371(99)00320-9} {\bibfield  {journal} {\bibinfo  {journal} {Physica A}\ }\textbf {\bibinfo {volume} {274}},\ \bibinfo {pages} {132} (\bibinfo {year} {1999})}\BibitemShut {NoStop}%
\bibitem [{\citenamefont {Neirotti}\ and\ \citenamefont {Caticha}(2024)}]{Neirotti2024}%
  \BibitemOpen
  \bibfield  {author} {\bibinfo {author} {\bibfnamefont {J.}~\bibnamefont {Neirotti}}\ and\ \bibinfo {author} {\bibfnamefont {N.}~\bibnamefont {Caticha}},\ }\bibfield  {title} {\bibinfo {title} {Legislative rebellions and impeachments in a neural network society},\ }\href {https://doi.org/10.1103/PhysRevE.110.054110} {\bibfield  {journal} {\bibinfo  {journal} {Phys. Rev. E}\ }\textbf {\bibinfo {volume} {110}},\ \bibinfo {pages} {054110} (\bibinfo {year} {2024})}\BibitemShut {NoStop}%
\bibitem [{\citenamefont {Klimek}\ \emph {et~al.}(2008)\citenamefont {Klimek}, \citenamefont {Lambiotte},\ and\ \citenamefont {Thurner}}]{Klimek_2008}%
  \BibitemOpen
  \bibfield  {author} {\bibinfo {author} {\bibfnamefont {P.}~\bibnamefont {Klimek}}, \bibinfo {author} {\bibfnamefont {R.}~\bibnamefont {Lambiotte}},\ and\ \bibinfo {author} {\bibfnamefont {S.}~\bibnamefont {Thurner}},\ }\bibfield  {title} {\bibinfo {title} {Opinion formation in laggard societies},\ }\href {https://doi.org/10.1209/0295-5075/82/28008} {\bibfield  {journal} {\bibinfo  {journal} {Europhys. Lett.}\ }\textbf {\bibinfo {volume} {82}},\ \bibinfo {pages} {28008} (\bibinfo {year} {2008})}\BibitemShut {NoStop}%
\bibitem [{\citenamefont {Castellano}\ \emph {et~al.}(2009)\citenamefont {Castellano}, \citenamefont {Fortunato},\ and\ \citenamefont {Loreto}}]{Castellano2009}%
  \BibitemOpen
  \bibfield  {author} {\bibinfo {author} {\bibfnamefont {C.}~\bibnamefont {Castellano}}, \bibinfo {author} {\bibfnamefont {S.}~\bibnamefont {Fortunato}},\ and\ \bibinfo {author} {\bibfnamefont {V.}~\bibnamefont {Loreto}},\ }\bibfield  {title} {\bibinfo {title} {Statistical physics of social dynamics},\ }\href {https://doi.org/10.1103/RevModPhys.81.591} {\bibfield  {journal} {\bibinfo  {journal} {Rev. Mod. Phys.}\ }\textbf {\bibinfo {volume} {81}},\ \bibinfo {pages} {591} (\bibinfo {year} {2009})}\BibitemShut {NoStop}%
\bibitem [{\citenamefont {Hegselmann}\ and\ \citenamefont {Krause}(2002)}]{Hegselmann2002}%
  \BibitemOpen
  \bibfield  {author} {\bibinfo {author} {\bibfnamefont {R.}~\bibnamefont {Hegselmann}}\ and\ \bibinfo {author} {\bibfnamefont {U.}~\bibnamefont {Krause}},\ }\bibfield  {title} {\bibinfo {title} {{Opinion Dynamics and Bounded Confidence Models, Analysis and Simulation}},\ }\href {https://ideas.repec.org/a/jas/jasssj/2002-5-2.html} {\bibfield  {journal} {\bibinfo  {journal} {J. Artif. Soc. Soc. Simul.}\ }\textbf {\bibinfo {volume} {5}},\ \bibinfo {pages} {1} (\bibinfo {year} {2002})}\BibitemShut {NoStop}%
\bibitem [{\citenamefont {Deffuant}\ \emph {et~al.}(2000)\citenamefont {Deffuant}, \citenamefont {Neau}, \citenamefont {Amblard},\ and\ \citenamefont {Weisbuch}}]{Deffuant2000}%
  \BibitemOpen
  \bibfield  {author} {\bibinfo {author} {\bibfnamefont {G.}~\bibnamefont {Deffuant}}, \bibinfo {author} {\bibfnamefont {D.}~\bibnamefont {Neau}}, \bibinfo {author} {\bibfnamefont {F.}~\bibnamefont {Amblard}},\ and\ \bibinfo {author} {\bibfnamefont {G.}~\bibnamefont {Weisbuch}},\ }\bibfield  {title} {\bibinfo {title} {Mixing beliefs among interacting agents},\ }\href {https://doi.org/10.1142/S0219525900000078} {\bibfield  {journal} {\bibinfo  {journal} {Adv. Complex Syst.}\ }\textbf {\bibinfo {volume} {03}},\ \bibinfo {pages} {87} (\bibinfo {year} {2000})}\BibitemShut {NoStop}%
\bibitem [{\citenamefont {Sznajd-Weron}\ and\ \citenamefont {Sznajd}(2000)}]{Sznajd2000}%
  \BibitemOpen
  \bibfield  {author} {\bibinfo {author} {\bibfnamefont {K.}~\bibnamefont {Sznajd-Weron}}\ and\ \bibinfo {author} {\bibfnamefont {J.}~\bibnamefont {Sznajd}},\ }\bibfield  {title} {\bibinfo {title} {Opinion evolution in closed community},\ }\href {https://doi.org/10.1142/S0129183100000936} {\bibfield  {journal} {\bibinfo  {journal} {Int. J. Mod. Phys. C}\ }\textbf {\bibinfo {volume} {11}},\ \bibinfo {pages} {1157} (\bibinfo {year} {2000})}\BibitemShut {NoStop}%
\bibitem [{\citenamefont {Axelrod}(1997)}]{Axelrod1997}%
  \BibitemOpen
  \bibfield  {author} {\bibinfo {author} {\bibfnamefont {R.}~\bibnamefont {Axelrod}},\ }\bibfield  {title} {\bibinfo {title} {The dissemination of culture: A model with local convergence and global polarization},\ }\href {http://www.jstor.org/stable/174371} {\bibfield  {journal} {\bibinfo  {journal} {J. Confl. Resolut.}\ }\textbf {\bibinfo {volume} {41}},\ \bibinfo {pages} {203} (\bibinfo {year} {1997})}\BibitemShut {NoStop}%
\bibitem [{\citenamefont {Liggett}(1999)}]{Liggett1999}%
  \BibitemOpen
  \bibfield  {author} {\bibinfo {author} {\bibfnamefont {T.~M.}\ \bibnamefont {Liggett}},\ }\bibinfo {title} {Voter models},\ in\ \href {https://doi.org/10.1007/978-3-662-03990-8_3} {\emph {\bibinfo {booktitle} {Stochastic Interacting Systems: Contact, Voter and Exclusion Processes}}}\ (\bibinfo  {publisher} {Springer Berlin Heidelberg},\ \bibinfo {address} {Berlin, Heidelberg},\ \bibinfo {year} {1999})\ pp.\ \bibinfo {pages} {139--208}\BibitemShut {NoStop}%
\bibitem [{\citenamefont {Sood}\ and\ \citenamefont {Redner}(2005)}]{Sood2005}%
  \BibitemOpen
  \bibfield  {author} {\bibinfo {author} {\bibfnamefont {V.}~\bibnamefont {Sood}}\ and\ \bibinfo {author} {\bibfnamefont {S.}~\bibnamefont {Redner}},\ }\bibfield  {title} {\bibinfo {title} {Voter model on heterogeneous graphs},\ }\href {https://doi.org/10.1103/PhysRevLett.94.178701} {\bibfield  {journal} {\bibinfo  {journal} {Phys. Rev. Lett.}\ }\textbf {\bibinfo {volume} {94}},\ \bibinfo {pages} {178701} (\bibinfo {year} {2005})}\BibitemShut {NoStop}%
\bibitem [{\citenamefont {Granovsky}\ and\ \citenamefont {Madras}(1995)}]{Granovsky1995}%
  \BibitemOpen
  \bibfield  {author} {\bibinfo {author} {\bibfnamefont {B.~L.}\ \bibnamefont {Granovsky}}\ and\ \bibinfo {author} {\bibfnamefont {N.}~\bibnamefont {Madras}},\ }\bibfield  {title} {\bibinfo {title} {The noisy voter model},\ }\href {https://doi.org/https://doi.org/10.1016/0304-4149(94)00035-R} {\bibfield  {journal} {\bibinfo  {journal} {Stoch. Process. Their Appl.}\ }\textbf {\bibinfo {volume} {55}},\ \bibinfo {pages} {23} (\bibinfo {year} {1995})}\BibitemShut {NoStop}%
\bibitem [{\citenamefont {Mobilia}(2003)}]{Mobilia2003}%
  \BibitemOpen
  \bibfield  {author} {\bibinfo {author} {\bibfnamefont {M.}~\bibnamefont {Mobilia}},\ }\bibfield  {title} {\bibinfo {title} {Does a single zealot affect an infinite group of voters?},\ }\href {https://doi.org/10.1103/PhysRevLett.91.028701} {\bibfield  {journal} {\bibinfo  {journal} {Phys. Rev. Lett.}\ }\textbf {\bibinfo {volume} {91}},\ \bibinfo {pages} {028701} (\bibinfo {year} {2003})}\BibitemShut {NoStop}%
\bibitem [{\citenamefont {Stark}\ \emph {et~al.}(2008)\citenamefont {Stark}, \citenamefont {Tessone},\ and\ \citenamefont {Schweitzer}}]{Stark2008}%
  \BibitemOpen
  \bibfield  {author} {\bibinfo {author} {\bibfnamefont {H.-U.}\ \bibnamefont {Stark}}, \bibinfo {author} {\bibfnamefont {C.~J.}\ \bibnamefont {Tessone}},\ and\ \bibinfo {author} {\bibfnamefont {F.}~\bibnamefont {Schweitzer}},\ }\bibfield  {title} {\bibinfo {title} {Decelerating microdynamics can accelerate macrodynamics in the voter model},\ }\href {https://doi.org/10.1103/PhysRevLett.101.018701} {\bibfield  {journal} {\bibinfo  {journal} {Phys. Rev. Lett.}\ }\textbf {\bibinfo {volume} {101}},\ \bibinfo {pages} {018701} (\bibinfo {year} {2008})}\BibitemShut {NoStop}%
\bibitem [{\citenamefont {Fern\'andez-Gracia}\ \emph {et~al.}(2014)\citenamefont {Fern\'andez-Gracia}, \citenamefont {Suchecki}, \citenamefont {Ramasco}, \citenamefont {San~Miguel},\ and\ \citenamefont {Egu\'{\i}luz}}]{Garcia2014}%
  \BibitemOpen
  \bibfield  {author} {\bibinfo {author} {\bibfnamefont {J.}~\bibnamefont {Fern\'andez-Gracia}}, \bibinfo {author} {\bibfnamefont {K.}~\bibnamefont {Suchecki}}, \bibinfo {author} {\bibfnamefont {J.~J.}\ \bibnamefont {Ramasco}}, \bibinfo {author} {\bibfnamefont {M.}~\bibnamefont {San~Miguel}},\ and\ \bibinfo {author} {\bibfnamefont {V.~M.}\ \bibnamefont {Egu\'{\i}luz}},\ }\bibfield  {title} {\bibinfo {title} {Is the voter model a model for voters?},\ }\href {https://doi.org/10.1103/PhysRevLett.112.158701} {\bibfield  {journal} {\bibinfo  {journal} {Phys. Rev. Lett.}\ }\textbf {\bibinfo {volume} {112}},\ \bibinfo {pages} {158701} (\bibinfo {year} {2014})}\BibitemShut {NoStop}%
\bibitem [{\citenamefont {Redner}(2019)}]{Redner2019}%
  \BibitemOpen
  \bibfield  {author} {\bibinfo {author} {\bibfnamefont {S.}~\bibnamefont {Redner}},\ }\bibfield  {title} {\bibinfo {title} {Reality-inspired voter models: A mini-review},\ }\href {https://doi.org/https://doi.org/10.1016/j.crhy.2019.05.004} {\bibfield  {journal} {\bibinfo  {journal} {C. R. Phys.}\ }\textbf {\bibinfo {volume} {20}},\ \bibinfo {pages} {275} (\bibinfo {year} {2019})}\BibitemShut {NoStop}%
\bibitem [{\citenamefont {McPherson}\ \emph {et~al.}(2001)\citenamefont {McPherson}, \citenamefont {Smith-Lovin},\ and\ \citenamefont {Cook}}]{McPherson2001}%
  \BibitemOpen
  \bibfield  {author} {\bibinfo {author} {\bibfnamefont {M.}~\bibnamefont {McPherson}}, \bibinfo {author} {\bibfnamefont {L.}~\bibnamefont {Smith-Lovin}},\ and\ \bibinfo {author} {\bibfnamefont {J.~M.}\ \bibnamefont {Cook}},\ }\bibfield  {title} {\bibinfo {title} {Birds of a feather: Homophily in social networks},\ }\href {https://doi.org/https://doi.org/10.1146/annurev.soc.27.1.415} {\bibfield  {journal} {\bibinfo  {journal} {Annu. Rev. Sociol.}\ }\textbf {\bibinfo {volume} {27}},\ \bibinfo {pages} {415} (\bibinfo {year} {2001})}\BibitemShut {NoStop}%
\bibitem [{\citenamefont {Heider}(1946)}]{Heider1946}%
  \BibitemOpen
  \bibfield  {author} {\bibinfo {author} {\bibfnamefont {F.}~\bibnamefont {Heider}},\ }\bibfield  {title} {\bibinfo {title} {Attitudes and cognitive organization},\ }\href {https://doi.org/10.1080/00223980.1946.9917275} {\bibfield  {journal} {\bibinfo  {journal} {J. Psychol.}\ }\textbf {\bibinfo {volume} {21}},\ \bibinfo {pages} {107} (\bibinfo {year} {1946})}\BibitemShut {NoStop}%
\bibitem [{\citenamefont {Marvel}\ \emph {et~al.}(2009)\citenamefont {Marvel}, \citenamefont {Strogatz},\ and\ \citenamefont {Kleinberg}}]{Marvel2009}%
  \BibitemOpen
  \bibfield  {author} {\bibinfo {author} {\bibfnamefont {S.~A.}\ \bibnamefont {Marvel}}, \bibinfo {author} {\bibfnamefont {S.~H.}\ \bibnamefont {Strogatz}},\ and\ \bibinfo {author} {\bibfnamefont {J.~M.}\ \bibnamefont {Kleinberg}},\ }\bibfield  {title} {\bibinfo {title} {Energy landscape of social balance},\ }\href {https://doi.org/10.1103/PhysRevLett.103.198701} {\bibfield  {journal} {\bibinfo  {journal} {Phys. Rev. Lett.}\ }\textbf {\bibinfo {volume} {103}},\ \bibinfo {pages} {198701} (\bibinfo {year} {2009})}\BibitemShut {NoStop}%
\bibitem [{\citenamefont {Minh~Pham}\ \emph {et~al.}(2020)\citenamefont {Minh~Pham}, \citenamefont {Kondor}, \citenamefont {Hanel},\ and\ \citenamefont {Thurner}}]{Pham2020}%
  \BibitemOpen
  \bibfield  {author} {\bibinfo {author} {\bibfnamefont {T.}~\bibnamefont {Minh~Pham}}, \bibinfo {author} {\bibfnamefont {I.}~\bibnamefont {Kondor}}, \bibinfo {author} {\bibfnamefont {R.}~\bibnamefont {Hanel}},\ and\ \bibinfo {author} {\bibfnamefont {S.}~\bibnamefont {Thurner}},\ }\bibfield  {title} {\bibinfo {title} {The effect of social balance on social fragmentation},\ }\href {https://doi.org/10.1098/rsif.2020.0752} {\bibfield  {journal} {\bibinfo  {journal} {J. R. Soc. Interface}\ }\textbf {\bibinfo {volume} {17}},\ \bibinfo {pages} {20200752} (\bibinfo {year} {2020})}\BibitemShut {NoStop}%
\bibitem [{\citenamefont {G\'orski}\ \emph {et~al.}(2020)\citenamefont {G\'orski}, \citenamefont {Bochenina}, \citenamefont {Ho\l{}yst},\ and\ \citenamefont {D'Souza}}]{Gorski2020}%
  \BibitemOpen
  \bibfield  {author} {\bibinfo {author} {\bibfnamefont {P.~J.}\ \bibnamefont {G\'orski}}, \bibinfo {author} {\bibfnamefont {K.}~\bibnamefont {Bochenina}}, \bibinfo {author} {\bibfnamefont {J.~A.}\ \bibnamefont {Ho\l{}yst}},\ and\ \bibinfo {author} {\bibfnamefont {R.~M.}\ \bibnamefont {D'Souza}},\ }\bibfield  {title} {\bibinfo {title} {Homophily based on few attributes can impede structural balance},\ }\href {https://doi.org/10.1103/PhysRevLett.125.078302} {\bibfield  {journal} {\bibinfo  {journal} {Phys. Rev. Lett.}\ }\textbf {\bibinfo {volume} {125}},\ \bibinfo {pages} {078302} (\bibinfo {year} {2020})}\BibitemShut {NoStop}%
\bibitem [{\citenamefont {Pham}\ \emph {et~al.}(2022)\citenamefont {Pham}, \citenamefont {Korbel}, \citenamefont {Hanel},\ and\ \citenamefont {Thurner}}]{Pham2022}%
  \BibitemOpen
  \bibfield  {author} {\bibinfo {author} {\bibfnamefont {T.~M.}\ \bibnamefont {Pham}}, \bibinfo {author} {\bibfnamefont {J.}~\bibnamefont {Korbel}}, \bibinfo {author} {\bibfnamefont {R.}~\bibnamefont {Hanel}},\ and\ \bibinfo {author} {\bibfnamefont {S.}~\bibnamefont {Thurner}},\ }\bibfield  {title} {\bibinfo {title} {Empirical social triad statistics can be explained with dyadic homophylic interactions},\ }\href {https://doi.org/10.1073/pnas.2121103119} {\bibfield  {journal} {\bibinfo  {journal} {PNAS}\ }\textbf {\bibinfo {volume} {119}},\ \bibinfo {pages} {e2121103119} (\bibinfo {year} {2022})}\BibitemShut {NoStop}%
\bibitem [{\citenamefont {Korbel}\ \emph {et~al.}(2023)\citenamefont {Korbel}, \citenamefont {Lindner}, \citenamefont {Pham}, \citenamefont {Hanel},\ and\ \citenamefont {Thurner}}]{Korbel2023}%
  \BibitemOpen
  \bibfield  {author} {\bibinfo {author} {\bibfnamefont {J.}~\bibnamefont {Korbel}}, \bibinfo {author} {\bibfnamefont {S.~D.}\ \bibnamefont {Lindner}}, \bibinfo {author} {\bibfnamefont {T.~M.}\ \bibnamefont {Pham}}, \bibinfo {author} {\bibfnamefont {R.}~\bibnamefont {Hanel}},\ and\ \bibinfo {author} {\bibfnamefont {S.}~\bibnamefont {Thurner}},\ }\bibfield  {title} {\bibinfo {title} {Homophily-based social group formation in a spin glass self-assembly framework},\ }\href {https://doi.org/10.1103/PhysRevLett.130.057401} {\bibfield  {journal} {\bibinfo  {journal} {Phys. Rev. Lett.}\ }\textbf {\bibinfo {volume} {130}},\ \bibinfo {pages} {057401} (\bibinfo {year} {2023})}\BibitemShut {NoStop}%
\bibitem [{\citenamefont {Galesic}\ \emph {et~al.}(2025)\citenamefont {Galesic}, \citenamefont {Olsson}, \citenamefont {Pham}, \citenamefont {Sorger},\ and\ \citenamefont {Thurner}}]{Galesic2025}%
  \BibitemOpen
  \bibfield  {author} {\bibinfo {author} {\bibfnamefont {M.}~\bibnamefont {Galesic}}, \bibinfo {author} {\bibfnamefont {H.}~\bibnamefont {Olsson}}, \bibinfo {author} {\bibfnamefont {T.~M.}\ \bibnamefont {Pham}}, \bibinfo {author} {\bibfnamefont {J.}~\bibnamefont {Sorger}},\ and\ \bibinfo {author} {\bibfnamefont {S.}~\bibnamefont {Thurner}},\ }\bibfield  {title} {\bibinfo {title} {Experimental evidence confirms that triadic social balance can be achieved through dyadic interactions},\ }\href {https://doi.org/10.1038/s44260-024-00022-y} {\bibfield  {journal} {\bibinfo  {journal} {npj Complexity}\ }\textbf {\bibinfo {volume} {2}},\ \bibinfo {pages} {1} (\bibinfo {year} {2025})}\BibitemShut {NoStop}%
\bibitem [{\citenamefont {Macy}\ \emph {et~al.}(2024)\citenamefont {Macy}, \citenamefont {Szymanski},\ and\ \citenamefont {Ho{\l}yst}}]{Macy2024}%
  \BibitemOpen
  \bibfield  {author} {\bibinfo {author} {\bibfnamefont {M.~W.}\ \bibnamefont {Macy}}, \bibinfo {author} {\bibfnamefont {B.~K.}\ \bibnamefont {Szymanski}},\ and\ \bibinfo {author} {\bibfnamefont {J.~A.}\ \bibnamefont {Ho{\l}yst}},\ }\bibfield  {title} {\bibinfo {title} {The ising model celebrates a century of interdisciplinary contributions},\ }\href {https://doi.org/10.1038/s44260-024-00012-0} {\bibfield  {journal} {\bibinfo  {journal} {npj Complexity}\ }\textbf {\bibinfo {volume} {1}},\ \bibinfo {pages} {10} (\bibinfo {year} {2024})}\BibitemShut {NoStop}%
\bibitem [{\citenamefont {Mullick}\ and\ \citenamefont {Sen}(2025)}]{Mullick2025}%
  \BibitemOpen
  \bibfield  {author} {\bibinfo {author} {\bibfnamefont {P.}~\bibnamefont {Mullick}}\ and\ \bibinfo {author} {\bibfnamefont {P.}~\bibnamefont {Sen}},\ }\bibfield  {title} {\bibinfo {title} {Sociophysics models inspired by the ising model},\ }\href {https://doi.org/10.1140/epjb/s10051-025-01053-7} {\bibfield  {journal} {\bibinfo  {journal} {The European Physical Journal B}\ }\textbf {\bibinfo {volume} {98}},\ \bibinfo {pages} {206} (\bibinfo {year} {2025})}\BibitemShut {NoStop}%
\bibitem [{\citenamefont {Starnini}\ \emph {et~al.}(2025)\citenamefont {Starnini}, \citenamefont {Baumann}, \citenamefont {Galla}, \citenamefont {Garcia}, \citenamefont {Iñiguez}, \citenamefont {Karsai}, \citenamefont {Lorenz},\ and\ \citenamefont {Sznajd-Weron}}]{starnini2025}%
  \BibitemOpen
  \bibfield  {author} {\bibinfo {author} {\bibfnamefont {M.}~\bibnamefont {Starnini}}, \bibinfo {author} {\bibfnamefont {F.}~\bibnamefont {Baumann}}, \bibinfo {author} {\bibfnamefont {T.}~\bibnamefont {Galla}}, \bibinfo {author} {\bibfnamefont {D.}~\bibnamefont {Garcia}}, \bibinfo {author} {\bibfnamefont {G.}~\bibnamefont {Iñiguez}}, \bibinfo {author} {\bibfnamefont {M.}~\bibnamefont {Karsai}}, \bibinfo {author} {\bibfnamefont {J.}~\bibnamefont {Lorenz}},\ and\ \bibinfo {author} {\bibfnamefont {K.}~\bibnamefont {Sznajd-Weron}},\ }\href {https://arxiv.org/abs/2507.11521} {\bibinfo {title} {Opinion dynamics: Statistical physics and beyond}} (\bibinfo {year} {2025}),\ \Eprint {https://arxiv.org/abs/2507.11521} {arXiv:2507.11521 [physics.soc-ph]} \BibitemShut {NoStop}%
\bibitem [{\citenamefont {Caldarelli}\ \emph {et~al.}(2025)\citenamefont {Caldarelli}, \citenamefont {Artime}, \citenamefont {Fischetti}, \citenamefont {Guarino}, \citenamefont {Nowak}, \citenamefont {Saracco}, \citenamefont {Holme},\ and\ \citenamefont {de~Domenico}}]{caldarelli2025}%
  \BibitemOpen
  \bibfield  {author} {\bibinfo {author} {\bibfnamefont {G.}~\bibnamefont {Caldarelli}}, \bibinfo {author} {\bibfnamefont {O.}~\bibnamefont {Artime}}, \bibinfo {author} {\bibfnamefont {G.}~\bibnamefont {Fischetti}}, \bibinfo {author} {\bibfnamefont {S.}~\bibnamefont {Guarino}}, \bibinfo {author} {\bibfnamefont {A.}~\bibnamefont {Nowak}}, \bibinfo {author} {\bibfnamefont {F.}~\bibnamefont {Saracco}}, \bibinfo {author} {\bibfnamefont {P.}~\bibnamefont {Holme}},\ and\ \bibinfo {author} {\bibfnamefont {M.}~\bibnamefont {de~Domenico}},\ }\href {https://arxiv.org/abs/2510.15053} {\bibinfo {title} {The physics of news, rumors, and opinions}} (\bibinfo {year} {2025}),\ \Eprint {https://arxiv.org/abs/2510.15053} {arXiv:2510.15053 [physics.soc-ph]} \BibitemShut {NoStop}%
\bibitem [{\citenamefont {Fortunato}\ and\ \citenamefont {Castellano}(2007)}]{Fortunato2007}%
  \BibitemOpen
  \bibfield  {author} {\bibinfo {author} {\bibfnamefont {S.}~\bibnamefont {Fortunato}}\ and\ \bibinfo {author} {\bibfnamefont {C.}~\bibnamefont {Castellano}},\ }\bibfield  {title} {\bibinfo {title} {Scaling and universality in proportional elections},\ }\href {https://doi.org/10.1103/PhysRevLett.99.138701} {\bibfield  {journal} {\bibinfo  {journal} {Phys. Rev. Lett.}\ }\textbf {\bibinfo {volume} {99}},\ \bibinfo {pages} {138701} (\bibinfo {year} {2007})}\BibitemShut {NoStop}%
\bibitem [{\citenamefont {Pal}\ \emph {et~al.}(2025)\citenamefont {Pal}, \citenamefont {Kumar},\ and\ \citenamefont {Santhanam}}]{Pal2025}%
  \BibitemOpen
  \bibfield  {author} {\bibinfo {author} {\bibfnamefont {R.}~\bibnamefont {Pal}}, \bibinfo {author} {\bibfnamefont {A.}~\bibnamefont {Kumar}},\ and\ \bibinfo {author} {\bibfnamefont {M.~S.}\ \bibnamefont {Santhanam}},\ }\bibfield  {title} {\bibinfo {title} {Universal statistics of competition in democratic elections},\ }\href {https://doi.org/10.1103/PhysRevLett.134.017401} {\bibfield  {journal} {\bibinfo  {journal} {Phys. Rev. Lett.}\ }\textbf {\bibinfo {volume} {134}},\ \bibinfo {pages} {017401} (\bibinfo {year} {2025})}\BibitemShut {NoStop}%
\bibitem [{\citenamefont {Tiwari}\ \emph {et~al.}(2021)\citenamefont {Tiwari}, \citenamefont {Yang},\ and\ \citenamefont {Sen}}]{Tiwari2021}%
  \BibitemOpen
  \bibfield  {author} {\bibinfo {author} {\bibfnamefont {M.}~\bibnamefont {Tiwari}}, \bibinfo {author} {\bibfnamefont {X.}~\bibnamefont {Yang}},\ and\ \bibinfo {author} {\bibfnamefont {S.}~\bibnamefont {Sen}},\ }\bibfield  {title} {\bibinfo {title} {Modeling the nonlinear effects of opinion kinematics in elections: A simple {I}sing model with random field-based study},\ }\href {https://doi.org/https://doi.org/10.1016/j.physa.2021.126287} {\bibfield  {journal} {\bibinfo  {journal} {Physica A}\ }\textbf {\bibinfo {volume} {582}},\ \bibinfo {pages} {126287} (\bibinfo {year} {2021})}\BibitemShut {NoStop}%
\bibitem [{\citenamefont {Thurner}\ \emph {et~al.}(2018)\citenamefont {Thurner}, \citenamefont {Hanel},\ and\ \citenamefont {Klimek}}]{thurner2018}%
  \BibitemOpen
  \bibfield  {author} {\bibinfo {author} {\bibfnamefont {S.}~\bibnamefont {Thurner}}, \bibinfo {author} {\bibfnamefont {R.}~\bibnamefont {Hanel}},\ and\ \bibinfo {author} {\bibfnamefont {P.}~\bibnamefont {Klimek}},\ }\href@noop {} {\emph {\bibinfo {title} {Introduction to the Theory of Complex Systems}}}\ (\bibinfo  {publisher} {Oxford University Press},\ \bibinfo {year} {2018})\BibitemShut {NoStop}%
\bibitem [{\citenamefont {Weidlich}(1991)}]{weidlich1991physics}%
  \BibitemOpen
  \bibfield  {author} {\bibinfo {author} {\bibfnamefont {W.}~\bibnamefont {Weidlich}},\ }\bibfield  {title} {\bibinfo {title} {Physics and social science—the approach of synergetics},\ }\href {https://doi.org/10.1016/0370-1573(91)90024-G} {\bibfield  {journal} {\bibinfo  {journal} {Phys. Rep.}\ }\textbf {\bibinfo {volume} {204}},\ \bibinfo {pages} {1} (\bibinfo {year} {1991})}\BibitemShut {NoStop}%
\bibitem [{\citenamefont {Weidlich}\ and\ \citenamefont {Haag}(2012)}]{weidlich2012concepts}%
  \BibitemOpen
  \bibfield  {author} {\bibinfo {author} {\bibfnamefont {W.}~\bibnamefont {Weidlich}}\ and\ \bibinfo {author} {\bibfnamefont {G.}~\bibnamefont {Haag}},\ }\href@noop {} {\emph {\bibinfo {title} {Concepts and models of a quantitative sociology: The dynamics of interacting populations}}},\ Vol.~\bibinfo {volume} {14}\ (\bibinfo  {publisher} {Springer Science \& Business Media},\ \bibinfo {year} {2012})\BibitemShut {NoStop}%
\bibitem [{\citenamefont {Imry}\ and\ \citenamefont {Ma}(1975)}]{Imry1975}%
  \BibitemOpen
  \bibfield  {author} {\bibinfo {author} {\bibfnamefont {Y.}~\bibnamefont {Imry}}\ and\ \bibinfo {author} {\bibfnamefont {S.-k.}\ \bibnamefont {Ma}},\ }\bibfield  {title} {\bibinfo {title} {Random-field instability of the ordered state of continuous symmetry},\ }\href {https://doi.org/10.1103/PhysRevLett.35.1399} {\bibfield  {journal} {\bibinfo  {journal} {Phys. Rev. Lett.}\ }\textbf {\bibinfo {volume} {35}},\ \bibinfo {pages} {1399} (\bibinfo {year} {1975})}\BibitemShut {NoStop}%
\bibitem [{\citenamefont {Bricmont}\ and\ \citenamefont {Kupiainen}(1987)}]{Bricmont1987}%
  \BibitemOpen
  \bibfield  {author} {\bibinfo {author} {\bibfnamefont {J.}~\bibnamefont {Bricmont}}\ and\ \bibinfo {author} {\bibfnamefont {A.}~\bibnamefont {Kupiainen}},\ }\bibfield  {title} {\bibinfo {title} {Lower critical dimension for the random-field {I}sing model},\ }\href {https://doi.org/10.1103/PhysRevLett.59.1829} {\bibfield  {journal} {\bibinfo  {journal} {Phys. Rev. Lett.}\ }\textbf {\bibinfo {volume} {59}},\ \bibinfo {pages} {1829} (\bibinfo {year} {1987})}\BibitemShut {NoStop}%
\bibitem [{\citenamefont {Fytas}\ \emph {et~al.}(2018)\citenamefont {Fytas}, \citenamefont {Mart{\'i}n-Mayor}, \citenamefont {Picco},\ and\ \citenamefont {Sourlas}}]{Fytas2018}%
  \BibitemOpen
  \bibfield  {author} {\bibinfo {author} {\bibfnamefont {N.~G.}\ \bibnamefont {Fytas}}, \bibinfo {author} {\bibfnamefont {V.}~\bibnamefont {Mart{\'i}n-Mayor}}, \bibinfo {author} {\bibfnamefont {M.}~\bibnamefont {Picco}},\ and\ \bibinfo {author} {\bibfnamefont {N.}~\bibnamefont {Sourlas}},\ }\bibfield  {title} {\bibinfo {title} {Review of recent developments in the random-field {I}sing model},\ }\href {https://doi.org/10.1007/s10955-018-1955-7} {\bibfield  {journal} {\bibinfo  {journal} {J. Stat. Phys.}\ }\textbf {\bibinfo {volume} {172}},\ \bibinfo {pages} {665} (\bibinfo {year} {2018})}\BibitemShut {NoStop}%
\bibitem [{\citenamefont {Hartmann}\ and\ \citenamefont {Nowak}(1999)}]{Hartmann1999}%
  \BibitemOpen
  \bibfield  {author} {\bibinfo {author} {\bibfnamefont {A.~K.}\ \bibnamefont {Hartmann}}\ and\ \bibinfo {author} {\bibfnamefont {U.}~\bibnamefont {Nowak}},\ }\bibfield  {title} {\bibinfo {title} {Universality in three dimensional random-field ground states},\ }\href {https://doi.org/10.1007/s100510050593} {\bibfield  {journal} {\bibinfo  {journal} {Eur. Phys. J. B}\ }\textbf {\bibinfo {volume} {7}},\ \bibinfo {pages} {105} (\bibinfo {year} {1999})}\BibitemShut {NoStop}%
\bibitem [{\citenamefont {Sinova}\ and\ \citenamefont {Canright}(2001)}]{Sinova2001}%
  \BibitemOpen
  \bibfield  {author} {\bibinfo {author} {\bibfnamefont {J.}~\bibnamefont {Sinova}}\ and\ \bibinfo {author} {\bibfnamefont {G.}~\bibnamefont {Canright}},\ }\bibfield  {title} {\bibinfo {title} {Nature and number of distinct phases in the random-field {I}sing model},\ }\href {https://doi.org/10.1103/PhysRevB.64.094402} {\bibfield  {journal} {\bibinfo  {journal} {Phys. Rev. B}\ }\textbf {\bibinfo {volume} {64}},\ \bibinfo {pages} {094402} (\bibinfo {year} {2001})}\BibitemShut {NoStop}%
\bibitem [{\citenamefont {Fytas}\ and\ \citenamefont {Malakis}(2008)}]{Fytas2008}%
  \BibitemOpen
  \bibfield  {author} {\bibinfo {author} {\bibfnamefont {N.~G.}\ \bibnamefont {Fytas}}\ and\ \bibinfo {author} {\bibfnamefont {A.}~\bibnamefont {Malakis}},\ }\bibfield  {title} {\bibinfo {title} {Phase diagram of the 3d bimodal random-field {I}sing model},\ }\href {https://doi.org/10.1140/epjb/e2008-00039-7} {\bibfield  {journal} {\bibinfo  {journal} {Eur. Phys. J. B}\ }\textbf {\bibinfo {volume} {61}},\ \bibinfo {pages} {111} (\bibinfo {year} {2008})}\BibitemShut {NoStop}%
\bibitem [{\citenamefont {Aharony}(1978)}]{Aharony1987}%
  \BibitemOpen
  \bibfield  {author} {\bibinfo {author} {\bibfnamefont {A.}~\bibnamefont {Aharony}},\ }\bibfield  {title} {\bibinfo {title} {Tricritical points in systems with random fields},\ }\href {https://doi.org/10.1103/PhysRevB.18.3318} {\bibfield  {journal} {\bibinfo  {journal} {Phys. Rev. B}\ }\textbf {\bibinfo {volume} {18}},\ \bibinfo {pages} {3318} (\bibinfo {year} {1978})}\BibitemShut {NoStop}%
\bibitem [{\citenamefont {Hadjiagapiou}(2010)}]{rfim2010}%
  \BibitemOpen
  \bibfield  {author} {\bibinfo {author} {\bibfnamefont {I.}~\bibnamefont {Hadjiagapiou}},\ }\bibfield  {title} {\bibinfo {title} {The random-field {I}sing model with asymmetric bimodal probability distribution},\ }\href {https://doi.org/https://doi.org/10.1016/j.physa.2010.05.033} {\bibfield  {journal} {\bibinfo  {journal} {Physica A}\ }\textbf {\bibinfo {volume} {389}},\ \bibinfo {pages} {3945} (\bibinfo {year} {2010})}\BibitemShut {NoStop}%
\bibitem [{\citenamefont {Galam}(1997)}]{Galam1997}%
  \BibitemOpen
  \bibfield  {author} {\bibinfo {author} {\bibfnamefont {S.}~\bibnamefont {Galam}},\ }\bibfield  {title} {\bibinfo {title} {Rational group decision making: A random field ising model at t = 0},\ }\href {https://doi.org/https://doi.org/10.1016/S0378-4371(96)00456-6} {\bibfield  {journal} {\bibinfo  {journal} {Physica A}\ }\textbf {\bibinfo {volume} {238}},\ \bibinfo {pages} {66} (\bibinfo {year} {1997})}\BibitemShut {NoStop}%
\bibitem [{\citenamefont {Bouchaud}(2013)}]{Bouchaud2013}%
  \BibitemOpen
  \bibfield  {author} {\bibinfo {author} {\bibfnamefont {J.-P.}\ \bibnamefont {Bouchaud}},\ }\bibfield  {title} {\bibinfo {title} {Crises and collective socio-economic phenomena: Simple models and challenges},\ }\href {https://doi.org/10.1007/s10955-012-0687-3} {\bibfield  {journal} {\bibinfo  {journal} {J. Stat. Phys.}\ }\textbf {\bibinfo {volume} {151}},\ \bibinfo {pages} {567} (\bibinfo {year} {2013})}\BibitemShut {NoStop}%
\bibitem [{\citenamefont {Anderson}\ and\ \citenamefont {Glomm}(1992)}]{Anderson1992}%
  \BibitemOpen
  \bibfield  {author} {\bibinfo {author} {\bibfnamefont {S.~P.}\ \bibnamefont {Anderson}}\ and\ \bibinfo {author} {\bibfnamefont {G.}~\bibnamefont {Glomm}},\ }\bibfield  {title} {\bibinfo {title} {Incumbency effects in political campaigns},\ }\href {https://doi.org/10.1007/BF00140768} {\bibfield  {journal} {\bibinfo  {journal} {Public Choice}\ }\textbf {\bibinfo {volume} {74}},\ \bibinfo {pages} {207} (\bibinfo {year} {1992})}\BibitemShut {NoStop}%
\bibitem [{\citenamefont {Fowler}(2018)}]{Fowler2018}%
  \BibitemOpen
  \bibfield  {author} {\bibinfo {author} {\bibfnamefont {A.}~\bibnamefont {Fowler}},\ }\bibfield  {title} {\bibinfo {title} {A {B}ayesian explanation for the effect of incumbency},\ }\href {https://doi.org/https://doi.org/10.1016/j.electstud.2018.03.005} {\bibfield  {journal} {\bibinfo  {journal} {Elect. Stud.}\ }\textbf {\bibinfo {volume} {53}},\ \bibinfo {pages} {66} (\bibinfo {year} {2018})}\BibitemShut {NoStop}%
\bibitem [{\citenamefont {Druckman}\ \emph {et~al.}(2020)\citenamefont {Druckman}, \citenamefont {Kifer},\ and\ \citenamefont {Parkin}}]{Druckman2020}%
  \BibitemOpen
  \bibfield  {author} {\bibinfo {author} {\bibfnamefont {J.~N.}\ \bibnamefont {Druckman}}, \bibinfo {author} {\bibfnamefont {M.~J.}\ \bibnamefont {Kifer}},\ and\ \bibinfo {author} {\bibfnamefont {M.}~\bibnamefont {Parkin}},\ }\bibfield  {title} {\bibinfo {title} {Campaign rhetoric and the incumbency advantage},\ }\href {https://doi.org/10.1177/1532673X18822314} {\bibfield  {journal} {\bibinfo  {journal} {Am. Politics Res.}\ }\textbf {\bibinfo {volume} {48}},\ \bibinfo {pages} {22} (\bibinfo {year} {2020})}\BibitemShut {NoStop}%
\bibitem [{\citenamefont {Hirsch}(2023)}]{Hirsch2023}%
  \BibitemOpen
  \bibfield  {author} {\bibinfo {author} {\bibfnamefont {A.~V.}\ \bibnamefont {Hirsch}},\ }\bibfield  {title} {\bibinfo {title} {Polarization and campaign spending in elections},\ }\href {https://doi.org/10.1086/722045} {\bibfield  {journal} {\bibinfo  {journal} {J. Politics}\ }\textbf {\bibinfo {volume} {85}},\ \bibinfo {pages} {240} (\bibinfo {year} {2023})}\BibitemShut {NoStop}%
\bibitem [{\citenamefont {Iyengar}\ \emph {et~al.}(2012)\citenamefont {Iyengar}, \citenamefont {Sood},\ and\ \citenamefont {Lelkes}}]{iyengar2012affect}%
  \BibitemOpen
  \bibfield  {author} {\bibinfo {author} {\bibfnamefont {S.}~\bibnamefont {Iyengar}}, \bibinfo {author} {\bibfnamefont {G.}~\bibnamefont {Sood}},\ and\ \bibinfo {author} {\bibfnamefont {Y.}~\bibnamefont {Lelkes}},\ }\bibfield  {title} {\bibinfo {title} {Affect, not ideology: A social identity perspective on polarization},\ }\href {https://doi.org/10.1093/poq/nfs038} {\bibfield  {journal} {\bibinfo  {journal} {Public Opin. Q.}\ }\textbf {\bibinfo {volume} {76}},\ \bibinfo {pages} {405} (\bibinfo {year} {2012})}\BibitemShut {NoStop}%
\bibitem [{\citenamefont {Lelkes}(2016)}]{lelkes2016mass}%
  \BibitemOpen
  \bibfield  {author} {\bibinfo {author} {\bibfnamefont {Y.}~\bibnamefont {Lelkes}},\ }\bibfield  {title} {\bibinfo {title} {Mass polarization: Manifestations and measurements},\ }\href {https://doi.org/10.1093/poq/nfw005} {\bibfield  {journal} {\bibinfo  {journal} {Public Opin. Q.}\ }\textbf {\bibinfo {volume} {80}},\ \bibinfo {pages} {392} (\bibinfo {year} {2016})}\BibitemShut {NoStop}%
\bibitem [{\citenamefont {Iyengar}\ \emph {et~al.}(2019)\citenamefont {Iyengar}, \citenamefont {Lelkes}, \citenamefont {Levendusky}, \citenamefont {Malhotra},\ and\ \citenamefont {Westwood}}]{iyengar2019origins}%
  \BibitemOpen
  \bibfield  {author} {\bibinfo {author} {\bibfnamefont {S.}~\bibnamefont {Iyengar}}, \bibinfo {author} {\bibfnamefont {Y.}~\bibnamefont {Lelkes}}, \bibinfo {author} {\bibfnamefont {M.}~\bibnamefont {Levendusky}}, \bibinfo {author} {\bibfnamefont {N.}~\bibnamefont {Malhotra}},\ and\ \bibinfo {author} {\bibfnamefont {S.~J.}\ \bibnamefont {Westwood}},\ }\bibfield  {title} {\bibinfo {title} {The origins and consequences of affective polarization in the united states},\ }\href {https://doi.org/10.1146/annurev-polisci-051117-073034} {\bibfield  {journal} {\bibinfo  {journal} {Annu. Rev. Poltical Sci.}\ }\textbf {\bibinfo {volume} {22}},\ \bibinfo {pages} {129} (\bibinfo {year} {2019})}\BibitemShut {NoStop}%
\bibitem [{FEC()}]{FEC.gov}%
  \BibitemOpen
  \href {https://www.fec.gov/} {\bibinfo {title} {https://www.fec.gov/}}\BibitemShut {NoStop}%
\bibitem [{\citenamefont {Bonica}(2024)}]{bonica2024dime}%
  \BibitemOpen
  \bibfield  {author} {\bibinfo {author} {\bibfnamefont {A.}~\bibnamefont {Bonica}},\ }\href@noop {} {\bibinfo {title} {Database on ideology, money in politics, and elections (dime)}},\ \bibinfo {howpublished} {\url{https://data.stanford.edu/dime}} (\bibinfo {year} {2024}),\ \bibinfo {note} {{S}tanford, CA}\BibitemShut {NoStop}%
\bibitem [{\citenamefont {Fouirnaies}\ and\ \citenamefont {Hall}(2014)}]{Fouirnaies2014}%
  \BibitemOpen
  \bibfield  {author} {\bibinfo {author} {\bibfnamefont {A.}~\bibnamefont {Fouirnaies}}\ and\ \bibinfo {author} {\bibfnamefont {A.~B.}\ \bibnamefont {Hall}},\ }\bibfield  {title} {\bibinfo {title} {The financial incumbency advantage: Causes and consequences},\ }\href {https://doi.org/10.1017/S0022381614000139} {\bibfield  {journal} {\bibinfo  {journal} {J. Politics}\ }\textbf {\bibinfo {volume} {76}},\ \bibinfo {pages} {711} (\bibinfo {year} {2014})}\BibitemShut {NoStop}%
\bibitem [{\citenamefont {{Pew Research Center}}(2017)}]{pewresearch_report}%
  \BibitemOpen
  \bibfield  {author} {\bibinfo {author} {\bibnamefont {{Pew Research Center}}},\ }\href {https://www.pewresearch.org/politics/2017/10/05/the-partisan-divide-on-political-values-grows-even-wider/} {\bibinfo {title} {The partisan divide on political values grows even wider}} (\bibinfo {year} {2017})\BibitemShut {NoStop}%
\bibitem [{\citenamefont {Jacobson}(2015)}]{Jacobson2015}%
  \BibitemOpen
  \bibfield  {author} {\bibinfo {author} {\bibfnamefont {G.~C.}\ \bibnamefont {Jacobson}},\ }\bibfield  {title} {\bibinfo {title} {It’s nothing personal: The decline of the incumbency advantage in us house elections},\ }\href {https://doi.org/10.1086/681670} {\bibfield  {journal} {\bibinfo  {journal} {J. Politics}\ }\textbf {\bibinfo {volume} {77}},\ \bibinfo {pages} {861} (\bibinfo {year} {2015})}\BibitemShut {NoStop}%
\bibitem [{\citenamefont {Stone}\ \emph {et~al.}(2010)\citenamefont {Stone}, \citenamefont {Fulton}, \citenamefont {Maestas},\ and\ \citenamefont {Maisel}}]{Stone2010}%
  \BibitemOpen
  \bibfield  {author} {\bibinfo {author} {\bibfnamefont {W.~J.}\ \bibnamefont {Stone}}, \bibinfo {author} {\bibfnamefont {S.~A.}\ \bibnamefont {Fulton}}, \bibinfo {author} {\bibfnamefont {C.~D.}\ \bibnamefont {Maestas}},\ and\ \bibinfo {author} {\bibfnamefont {L.~S.}\ \bibnamefont {Maisel}},\ }\bibfield  {title} {\bibinfo {title} {Incumbency reconsidered: Prospects, strategic retirement, and incumbent quality in u.s. house elections},\ }\href {https://doi.org/10.1017/S0022381609990557} {\bibfield  {journal} {\bibinfo  {journal} {J. Politics}\ }\textbf {\bibinfo {volume} {72}},\ \bibinfo {pages} {178} (\bibinfo {year} {2010})}\BibitemShut {NoStop}%
\bibitem [{\citenamefont {Blankschtein}\ \emph {et~al.}(1984)\citenamefont {Blankschtein}, \citenamefont {Shapir},\ and\ \citenamefont {Aharony}}]{Blankschtein1984}%
  \BibitemOpen
  \bibfield  {author} {\bibinfo {author} {\bibfnamefont {D.}~\bibnamefont {Blankschtein}}, \bibinfo {author} {\bibfnamefont {Y.}~\bibnamefont {Shapir}},\ and\ \bibinfo {author} {\bibfnamefont {A.}~\bibnamefont {Aharony}},\ }\bibfield  {title} {\bibinfo {title} {Potts models in random fields},\ }\href {https://doi.org/10.1103/PhysRevB.29.1263} {\bibfield  {journal} {\bibinfo  {journal} {Phys. Rev. B}\ }\textbf {\bibinfo {volume} {29}},\ \bibinfo {pages} {1263} (\bibinfo {year} {1984})}\BibitemShut {NoStop}%
\bibitem [{\citenamefont {Jacobson}(1978)}]{Jacobson1978}%
  \BibitemOpen
  \bibfield  {author} {\bibinfo {author} {\bibfnamefont {G.~C.}\ \bibnamefont {Jacobson}},\ }\bibfield  {title} {\bibinfo {title} {The effects of campaign spending in congressional elections},\ }\href {https://doi.org/10.2307/1954105} {\bibfield  {journal} {\bibinfo  {journal} {Am. Political Sci. Rev.}\ }\textbf {\bibinfo {volume} {72}},\ \bibinfo {pages} {469–491} (\bibinfo {year} {1978})}\BibitemShut {NoStop}%
\bibitem [{\citenamefont {Jacobson}(1990)}]{Jacobson1990}%
  \BibitemOpen
  \bibfield  {author} {\bibinfo {author} {\bibfnamefont {G.~C.}\ \bibnamefont {Jacobson}},\ }\bibfield  {title} {\bibinfo {title} {The effects of campaign spending in house elections: New evidence for old arguments},\ }\href {http://www.jstor.org/stable/2111450} {\bibfield  {journal} {\bibinfo  {journal} {Am. J. Political Sci.}\ }\textbf {\bibinfo {volume} {34}},\ \bibinfo {pages} {334} (\bibinfo {year} {1990})}\BibitemShut {NoStop}%
\bibitem [{\citenamefont {Erikson}\ and\ \citenamefont {Palfrey}(2000)}]{Erikson2000}%
  \BibitemOpen
  \bibfield  {author} {\bibinfo {author} {\bibfnamefont {R.~S.}\ \bibnamefont {Erikson}}\ and\ \bibinfo {author} {\bibfnamefont {T.~R.}\ \bibnamefont {Palfrey}},\ }\bibfield  {title} {\bibinfo {title} {Equilibria in campaign spending games: Theory and data},\ }\href {https://doi.org/10.2307/2585833} {\bibfield  {journal} {\bibinfo  {journal} {Am. Political Sci. Rev.}\ }\textbf {\bibinfo {volume} {94}},\ \bibinfo {pages} {595–609} (\bibinfo {year} {2000})}\BibitemShut {NoStop}%
\bibitem [{\citenamefont {Schnakenberg}\ and\ \citenamefont {Turner}(2021)}]{Schnakenberg2021}%
  \BibitemOpen
  \bibfield  {author} {\bibinfo {author} {\bibfnamefont {K.~E.}\ \bibnamefont {Schnakenberg}}\ and\ \bibinfo {author} {\bibfnamefont {I.~R.}\ \bibnamefont {Turner}},\ }\bibfield  {title} {\bibinfo {title} {Helping friends or influencing foes: Electoral and policy effects of campaign finance contributions},\ }\href {https://doi.org/https://doi.org/10.1111/ajps.12534} {\bibfield  {journal} {\bibinfo  {journal} {Am. J. Political Sci.}\ }\textbf {\bibinfo {volume} {65}},\ \bibinfo {pages} {88} (\bibinfo {year} {2021})}\BibitemShut {NoStop}%
\bibitem [{\citenamefont {Avis}\ \emph {et~al.}(2022)\citenamefont {Avis}, \citenamefont {Ferraz}, \citenamefont {Finan},\ and\ \citenamefont {Varjão}}]{Avis2022}%
  \BibitemOpen
  \bibfield  {author} {\bibinfo {author} {\bibfnamefont {E.}~\bibnamefont {Avis}}, \bibinfo {author} {\bibfnamefont {C.}~\bibnamefont {Ferraz}}, \bibinfo {author} {\bibfnamefont {F.}~\bibnamefont {Finan}},\ and\ \bibinfo {author} {\bibfnamefont {C.}~\bibnamefont {Varjão}},\ }\bibfield  {title} {\bibinfo {title} {Money and politics: The effects of campaign spending limits on political entry and competition},\ }\href {https://doi.org/10.1257/app.20200296} {\bibfield  {journal} {\bibinfo  {journal} {Am. Econ. J. Appl. Econ.}\ }\textbf {\bibinfo {volume} {14}},\ \bibinfo {pages} {167–99} (\bibinfo {year} {2022})}\BibitemShut {NoStop}%
\bibitem [{\citenamefont {Lu}\ \emph {et~al.}(2019)\citenamefont {Lu}, \citenamefont {Gao},\ and\ \citenamefont {Szymanski}}]{luEvolutionPolarizationLegislative2019}%
  \BibitemOpen
  \bibfield  {author} {\bibinfo {author} {\bibfnamefont {X.}~\bibnamefont {Lu}}, \bibinfo {author} {\bibfnamefont {J.}~\bibnamefont {Gao}},\ and\ \bibinfo {author} {\bibfnamefont {B.~K.}\ \bibnamefont {Szymanski}},\ }\bibfield  {title} {\bibinfo {title} {The evolution of polarization in the legislative branch of government},\ }\href {https://doi.org/10.1098/rsif.2019.0010} {\bibfield  {journal} {\bibinfo  {journal} {J. R. Soc. Interface}\ }\textbf {\bibinfo {volume} {16}},\ \bibinfo {pages} {20190010} (\bibinfo {year} {2019})}\BibitemShut {NoStop}%
\bibitem [{\citenamefont {Thurner}\ \emph {et~al.}(2025)\citenamefont {Thurner}, \citenamefont {Hofer},\ and\ \citenamefont {Korbel}}]{Thurner2025}%
  \BibitemOpen
  \bibfield  {author} {\bibinfo {author} {\bibfnamefont {S.}~\bibnamefont {Thurner}}, \bibinfo {author} {\bibfnamefont {M.}~\bibnamefont {Hofer}},\ and\ \bibinfo {author} {\bibfnamefont {J.}~\bibnamefont {Korbel}},\ }\bibfield  {title} {\bibinfo {title} {Why more social interactions lead to more polarization in societies},\ }\href {https://doi.org/10.1073/pnas.2517530122} {\bibfield  {journal} {\bibinfo  {journal} {PNAS}\ }\textbf {\bibinfo {volume} {122}},\ \bibinfo {pages} {e2517530122} (\bibinfo {year} {2025})}\BibitemShut {NoStop}%
\bibitem [{\citenamefont {Pham}\ \emph {et~al.}(2025)\citenamefont {Pham}, \citenamefont {Redner}, \citenamefont {Waldorp}, \citenamefont {Armas},\ and\ \citenamefont {van~der Maas}}]{Pham2025}%
  \BibitemOpen
  \bibfield  {author} {\bibinfo {author} {\bibfnamefont {T.}~\bibnamefont {Pham}}, \bibinfo {author} {\bibfnamefont {S.}~\bibnamefont {Redner}}, \bibinfo {author} {\bibfnamefont {L.}~\bibnamefont {Waldorp}}, \bibinfo {author} {\bibfnamefont {J.}~\bibnamefont {Armas}},\ and\ \bibinfo {author} {\bibfnamefont {H.~L.~J.}\ \bibnamefont {van~der Maas}},\ }\href {https://arxiv.org/abs/2503.24098} {\bibinfo {title} {Polarisation in increasingly connected societies}} (\bibinfo {year} {2025}),\ \Eprint {https://arxiv.org/abs/2503.24098} {arXiv:2503.24098 [physics.soc-ph]} \BibitemShut {NoStop}%
\bibitem [{\citenamefont {Golan}\ \emph {et~al.}(2000)\citenamefont {Golan}, \citenamefont {Karp},\ and\ \citenamefont {Perloff}}]{Golan2000}%
  \BibitemOpen
  \bibfield  {author} {\bibinfo {author} {\bibfnamefont {A.}~\bibnamefont {Golan}}, \bibinfo {author} {\bibfnamefont {L.~S.}\ \bibnamefont {Karp}},\ and\ \bibinfo {author} {\bibfnamefont {J.~M.}\ \bibnamefont {Perloff}},\ }\bibfield  {title} {\bibinfo {title} {Estimating {C}oke's and {P}epsi's price and advertising strategies},\ }\href {https://doi.org/10.1080/07350015.2000.10524880} {\bibfield  {journal} {\bibinfo  {journal} {J. Bus. Econ. Stat.}\ }\textbf {\bibinfo {volume} {18}},\ \bibinfo {pages} {398} (\bibinfo {year} {2000})}\BibitemShut {NoStop}%
\end{thebibliography}%

\clearpage

\onecolumngrid

\section*{End matter}

\clearpage
\onecolumngrid

\section*{SUPPLEMENTAL MATERIAL}
\subsection*{Detailed derivation of the self-consistency equation}
\noindent We start with the Hamiltonian
\begin{equation*}
 H(s_1,\dots,s_N) = -J \sum_{i < j} A_{ij}s_i s_j - \sum_i h_i s_i\, .  
\end{equation*}
In order to decouple the Hamiltonian into the Hamiltonian for a single individual, we use the two approximations. The first one is the \emph{configuration model approximation}, where we assume that the adjacency matrix of a random network can be approximated as $A_{ij} \approx \frac{k_i k_j}
{N\langle k \rangle}$, where $k_i$ is the degree (i.e., connectivity) of the node $i$ and $\langle k \rangle$ is the average degree. The second approximation we use is the \emph{mean-field approximation}. Expressing the spin in terms of its average value as  $s_i = \langle s_i \rangle + \delta s_i$
enables us to omit the term quadratic in fluctuations. By denoting the average magnetization as $m = \langle s_i\rangle$, we can rewrite the Hamiltonian as 
\begin{eqnarray*}
H(s_1,\dots,s_N) &\approx&  \frac{J m^2 N \langle k\rangle}{2} - J \sum_i \langle k \rangle\, m s_i - \sum_i h_i s_i \, .
\end{eqnarray*}
The first term can be omitted from the Hamiltonian, as it is an additive constant of the energy and will be canceled when the equilibrium distribution is calculated. Thus, the mean-field Hamiltonian can be expressed as
\begin{equation*}
H^{MF}(s_1,\dots,s_N) = -  \sum_i (\tilde{J} m  + h_i) s_i
\end{equation*}
where $\tilde{J} = J \langle k \rangle$. Note that $h_i$ is a random variable with the distribution 
\begin{equation*}
p(h_i) = p \delta(h_i - h^+) + (1-p) \delta(h_i + h^{-})\, .
\end{equation*}
The equilibrium distribution is therefore
\begin{equation*}
p(s|h^\pm) = \exp\left(-\beta (\tilde{J} m \pm h^\pm)s\right)/Z^{\pm}
\end{equation*}
where $\beta = \frac{1}{k T}$ is the inverse temperature (we set $k=1$), and 
\begin{equation*}
Z^{\pm} = 2 \cosh\left(\beta (\tilde{J} m \pm h^{\pm})\right)
\end{equation*} is the partition function. The average magnetization of spins coupled to external field $\pm h^{\pm}$ is 
\begin{equation*}
m^{\pm} = \langle s \rangle^{\pm} = \sum_{s= \pm 1} s p(s|h^\pm)   = \tanh\left(\beta(\tilde{J} m \pm h^{\pm})\right)\, .
\end{equation*}
The total magnetization can be expressed as 
\begin{eqnarray*}
m &=& \langle m \rangle_h = p m^+ + (1-p) m^- \nonumber\\
&=& p \tanh\left(\beta(\tilde{J} m + h^{+})\right) + (1-p) \tanh\left(\beta(\tilde{J}m - h^-)\right)\, . 
\end{eqnarray*}
\subsection*{Derivation of the critical curve and tricritical point for symmetric case}
Let us now focus on the symmetric case, i.e., when $p=\frac{1}{2}$ and $h^+ = h^- \equiv h$. We investigate how the phase diagram depends on the strength of the field and the temperature. Without loss of generality, we consider that $\tilde{J}=1$. To this end, we expand the right-hand side of the self-consistency equation around $m=0$ and get

\begin{equation*}
m = \frac{1-\xi^2}{T} m+ \frac{4 \xi^2-3\xi^4-1}{3T^3} m^3 + \mathcal{O}(m^5)    
\end{equation*}
where $\xi \equiv \tanh(h/T)$.
This cubic equation has three solutions: one trivial $m^0=0$ and two non-trivial solutions 
\begin{equation*}
m^{\pm} = \pm \frac{\sqrt{3(1-T-\xi^2})}{\sqrt{\frac{T^2+3 \xi^4-4\xi^2}{T^2}}} \, .
\end{equation*}
By comparing when the non-trivial solutions become trivial, i.e., $m^0 = m^{\pm}$, we obtain the critical curve 
\begin{equation*}
h_c = T \textrm{arctanh}(\sqrt{1-T})\, .
\end{equation*}
Finally, by plugging the critical curve into the self-consistency equation, we can determine the order of the phase transition from the sign of the third-order coefficient. The coefficient is along the critical curve equal to $\frac{2-3 T}{3 T^2}$ so the critical point where the phase transition changes its order is
\begin{equation*}
T^* = \frac{2}{3}\, , \qquad h^*=\frac{2}{3} \, \textrm{arctanh}\left(\frac{1}{\sqrt{3}}\right) \approx 0.439\, .    
\end{equation*}



\subsection*{Derivation of the critical curve and tricritical point for the asymmetric case}
Let us now focus on the general case when $p \neq \frac{1}{2}$. We take the condition for $m=0$, which is
\begin{equation*}
p \tanh(\beta h^+) = (1-p) \tanh(\beta h^-)
\end{equation*}
and expand the self-consistency equation 
\begin{equation*}
m = p \tanh\left(\beta(\tilde{J} m + h^{+})\right) + (1-p) \tanh\left(\beta(\tilde{J}m - h^-)\right)\, .
\end{equation*}
around $m=0$ while keeping the dependence between $h^+$ and $h^-$ determined from the condition on $m=0$ above.  We use the Taylor expansion of $\tanh(\beta(m+h))$ which is
\begin{eqnarray*}
\tanh(\beta(m\pm h^\pm)) = \pm \tanh(\beta h^\pm)
+ \beta m \left(1-\tanh^2(\beta h^\pm)\right) 
\mp \beta^2 m^2 \tanh(\beta h^\pm) (1-\tanh^2(\beta h^\pm)) + \mathcal{O}(m^3)\, .
\end{eqnarray*}
By denoting $\xi^\pm = \tanh(h^\pm/T)$, one can rewrite the self-consistency equation as
\begin{eqnarray*}
m = p \left[\xi^+ +  m \frac{1-(\xi^+)^2}{T} - m^2 \frac{\xi^+ (1-(\xi^+)^2)}{T^2}\right]
+ (1-p) \left[-\xi^- + m \frac{1-(\xi^-)^2}{T} + m^2 \frac{\xi^-(1-(\xi^-)^2)}{T^2}\right] + \mathcal{O}(m^3)\, .
\end{eqnarray*}
Since $p \, \xi^+ - (1-p) \, \xi^-=0$, the constant term is zero. Therefore, we can rearrange the terms as
\begin{eqnarray*}
\left[\frac{T - p(1-(\xi^+)^2)- (1-p) (1-(\xi^-)^2)}{T}\right]m 
+ \left[ \frac{p \xi^+ (1-(\xi^+)^2) - (1-p) \xi^- (1-(\xi^-)^2)}{T^2}\right]m^2 =0\, .
\end{eqnarray*}
The solution is therefore either $m_0 = 0$ or 
\begin{equation*}
m_1 = - T\frac{T - p(1-(\xi^+)^2)- (1-p) (1-(\xi^-)^2)}{p \xi^+ (1-(\xi^+)^2) - (1-p) \xi^- (1-(\xi^-)^2)}\, . 
\end{equation*}
Now, the critical point is given by the condition $m_1=m_0 \equiv 0$, which is equivalent to 
\begin{equation*}
    T = p(1-(\xi^+_c)^2)+ (1-p) (1-(\xi^-_c)^2)\, . 
\end{equation*}
By plugging in from the condition $\xi^- = \frac{p}{1-p} \xi^+$, we obtain
\begin{eqnarray*}
 T = p (1-(\xi^+_c)^2) + (1-p) \left[1- \left(\frac{p}{1-p}\right)^2 (\xi^+_c)^2\right] 
 = 1 - \frac{p}{1-p} (\xi^+_c)^2\, .
\end{eqnarray*}
By plugging back for the $\xi_c^+= \tanh(h^+_c/T)$, we can express $h_c^+$ on $T$ as
\begin{equation*}
h_c^+ =  T \,\textrm{arctanh}  \sqrt{(1-T)\, \frac{1-p}{p}}\, .
\end{equation*}
Similarly, by expressing $h_c^-$ from the condition, we get that
\begin{equation*}
h_c^- = T \, \textrm{arctanh}  \sqrt{(1-T)\, \frac{p}{1-p}}\, . 
\end{equation*}
Interestingly, for $p=\frac{1}{2}$, both equations boild down to $h_c^+ = h_c^- \equiv h_c$.

\subsection*{Alternative derivation of the self-consistency equation from the Weidlich master-equation model}
In this section, we show that the mean-field description used in the main text can alternatively be obtained from a master-equation approach to opinion dynamics, following the sociodynamics framework introduced by Weidlich and Haag. This demonstrates that the polarization transition does not rely on the Hamiltonian formulation with spatial interactions, but emerges generically from stochastic opinion switching driven by individual preferences and social adaptation.

Let us consider a population of $N$ voters with binary opinions $s_i \in \pm 1$. Let us define the number of voters with a positive opinion as $n$, and the fraction of voters with opinion $s = +1$ as $x = n/N \in [0,1]$. The natural connection to the magnetization defined in the main paper is $m=2x-1$. The system undergoes a Markovian evolution described by a master equation, fully characterized by the transition rates $W_{+1 \, \rightarrow \, -1}(n) \equiv W_{+ -} (n)$ and $W_{-1 \, \rightarrow \, +1}(n) \equiv W_{-+}(n)$. The master equation can therefore be expressed as

\begin{eqnarray*} 
 \frac{\mathrm{d} P(n)}{\mathrm{d} t} &=& (N-n+1)W_{-+}(n-1) P(n-1,t)\\
 &+&  (n+1)W_{+-}(n+1) P(n+1,t)\\
 &-&[(N-n) W_{-+}(n) + n W_{+-}(n)] P(n,t)\, .
\end{eqnarray*}

In the classic formulation of sociodynamics, the individual opinion changes are governed by two conceptually distinct mechanisms. First, individuals possess intrinsic preferences that are independent of the current social configuration and may reflect long-standing inclinations, prior beliefs, or external information. In the election setting, this corresponds to the election campaign they follow

Second, individuals exhibit an adaptive response to the prevailing opinion, whereby the propensity to adopt a given opinion increases with its share of the population, reflecting conformity or social pressure. This mechanism is somewhat similar to the homophily, although it does not necessarily require the assumption on the spatial distribution of interactions. These preference and adaptation tendencies act simultaneously at the individual level and may have independent strengths. The particular choice that is widely used in sociodynamics literature is

\begin{eqnarray*}
 W_{-+}(n) &=& \nu \exp(\theta + K x )\, ,\\
 W_{+-}(n) &=&  \nu \exp(-(\theta+ Kx))\, ,
\end{eqnarray*}
where $\theta$ quantifies the former mechanism, while $K x$ the latter one. In our setting, $\theta$ represents the strength of the election campaign. Thus, in our scenario, we divide the population into two subpopulations, one following the campaign of $s=+1$ of size $N^+$ and the other following the campaign of $s=-1$ of size $N^-$. We denote the number of individuals following the first campaign with a positive opinion as $n^+$ and analogously $n^-$.  The campaign intensities are $\theta^+$ and $-\theta^-$, and the transition rates for gives subpopulations are therefore 
\begin{eqnarray*}
 W_{-+}^+(n^+) &=& \nu \exp(\theta^+ + K x )\, ,\\
 W_{+-}^+(n^+) &=&  \nu \exp(-(\theta^++ Kx))\, ,\\
 W_{-+}^-(n^-) &=& \nu \exp(-\theta^- + K x )\, ,\\
 W_{+-}^-(n^-) &=&  \nu \exp(-(-\theta^-+ Kx))\, .
\end{eqnarray*}
Here, individuals pursue their own campaigns but still seek to align with the majority of the population, regardless of the campaign. This is reflected by the term $K x$.
By calculating the first moment of the probability distribution from the master equation for each population in the case of $N\gg 1$, we obtain
\begin{eqnarray*}
    \dot{x}^{+} &=& (1-x^+) \nu e^{\theta^+ + Kx} - x^+ \nu e^{-(\theta^+ + Kx)}\, ,\\
    \dot{x}^- &=& (1-x^-) \nu e^{-\theta^- + Kx} - x^- \nu e^{-(-\theta^-+Kx)}\, .
\end{eqnarray*}
At the stationary point, we obtain, after a straightforward calculation
$$x^\pm = \frac{1}{2}\left[1 + \tanh\!\big(Kx \pm \theta^\pm\big)\right]\, .$$
By defining $p=N^+/N$, we can rewrite normalization as
$x = p x^+ + (1-p) x^-$, and therefore we obtain
$$x = p \, \frac{1}{2} \left[1 + \tanh(Kx + \theta^+)\right]+ (1-p) \, \frac{1}{2} \left[1 + \tanh(Kx - \theta^-)\right]\, .$$
Using the relation $m = 2x-1$, the term $K x$ can be rewritten as $\frac{K m}{2} + \frac{K}{2}$ where the constant contribution is absorbed into a redefinition of the effective fields. Thus, by choosing the following transformation
\begin{eqnarray*}
m &=& 2x-1\, ,\\
2\beta \tilde{J} &=& K\, ,\\
\beta h^+ &=& \theta^+ + \frac{K}{2}\, ,\\
\beta h^- &=& \theta^- - \frac{K}{2}\, ,
\end{eqnarray*}
we obtain the equation for $m$
$$m = p \tanh\left(\beta(\tilde{J} m + h^{+})\right) + (1-p) \tanh\left(\beta(\tilde{J}m - h^-)\right)$$ which is exactly the self-consistency equation from the main text. 

\subsection*{Classification model} Here, we describe the classification models used in the main text. The classification model is directly based on the results of the Random Field Ising model. Without loss of generality, we assign $h^+=h^{REP}$ as the spending of the Republican party candidate, $h^-=h^{DEM}$ as the spending of the Democratic party candidate. The prediction of the classification model, based on the magnetization $m$ goes as follows:
\begin{itemize}
    \item If $(h^{DEM},h^{REP})$ lie in the hysteresis region, then the model predicts the incumbent as the winner.
    \item If $(h^{DEM},h^{REP})$ lie outside of the hysteresis region, or if there is an open seat (i.e., the incumbent does not run as a candidate), a Republican wins if $m>0$, and a Democrat wins if $m<0$.
\end{itemize}
Since, for the temperature $T \geq 1$, we observe no hysteresis in the region, only the second condition applies. Specifically, when $p=0.5$, the condition on the sign of magnetization $m$ boils down to the condition whether $h^{REP} > h^{DEM}$ (corresponding to $m>0$) or the other way around. In this case, the classification model does not depend on temperature $T$ (when $T\geq 1)$. We call this model the \emph{null model}. This model catches the intuitive idea that in the case of equal campaign coverage, the candidate who spends more money on the campaign wins the election. 

\subsection*{Model accuracy} In order to measure the performance of the classification model, we use the model \emph{accuracy}. The confusion table between the predicted classification and the actual classification (here, the prediction is that a Republican candidate wins an election) is then defined in Tab. \ref{tab:ct}. 
The accuracy is defined as
$$ACC = \frac{n_{TP}+n_{TN}}{n_{TP} + n_{TN} + n_{FP} + n_{FN}}\, .$$

\begin{table}[h]
    \centering
\begin{tabular}{|c|c|c|}
    \hline
     predicted/actual & &  \\
     classification & positive & negative\\
     \hline 
    positive &  True positive ($n_{TP}$) & False negative ($n_{FN}$)  \\
    negative &   False positive ($n_{FP}$) & True negative ($n_{TN}$)\\
    \hline
\end{tabular}
\caption{Confusion table of the classification model.}
\label{tab:ct}
\end{table}

\subsection*{McNemar test} The McNemar test is used to demonstrate whether one of the two classification models used on a given data set is better than the other. For each observation, a classification model gives a predicted classification, which is compared with the actual classification. For example, in the election races, the classification model predicts the winner of the election based on the campaign spending and incumbency (see the second above), which is then compared with the actual election result. For two classification models $M_1$,$M_2$, the contingency table between correctly and incorrectly classified observations can be written as shown in Tab. \ref{tab:MN}.
\begin{table}[h]
    \centering
\begin{tabular}{|c|c|c|}
\hline
     $M_1$/$M_2$ & $M_2$ correct & $M_2$ wrong\\
     \hline 
    $M_1$ correct &  $n_{11}$ & $n_{12}$  \\
    $M_1$ wrong &   $n_{21}$ & $n_{22}$ \\
    \hline
\end{tabular}
\caption{Contingency table for the McNemar test.}
\label{tab:MN}
\end{table}

The null hypothesis is that both marginals are the same, and therefore the probability that the first model is correct and the second model is wrong is the same as that the first model is wrong and the second model is correct.
The McNemar test statistic is 
$$\chi^2 = \frac{(n_{12}-n_{21})^2}{n_{12}+n_{21}}\, .$$
Under the assumption of the null hypothesis, and for large enough $n_{12}$ and $n_{21}$, the statistic follows a $\chi^2$ distribution with one degree of freedom. We can therefore reject the null hypothesis if the observed statistic is significant (i.e., the $p$-value is smaller than the desired statistical level).

\begin{figure*}[h!]
\includegraphics[width=0.35\linewidth]{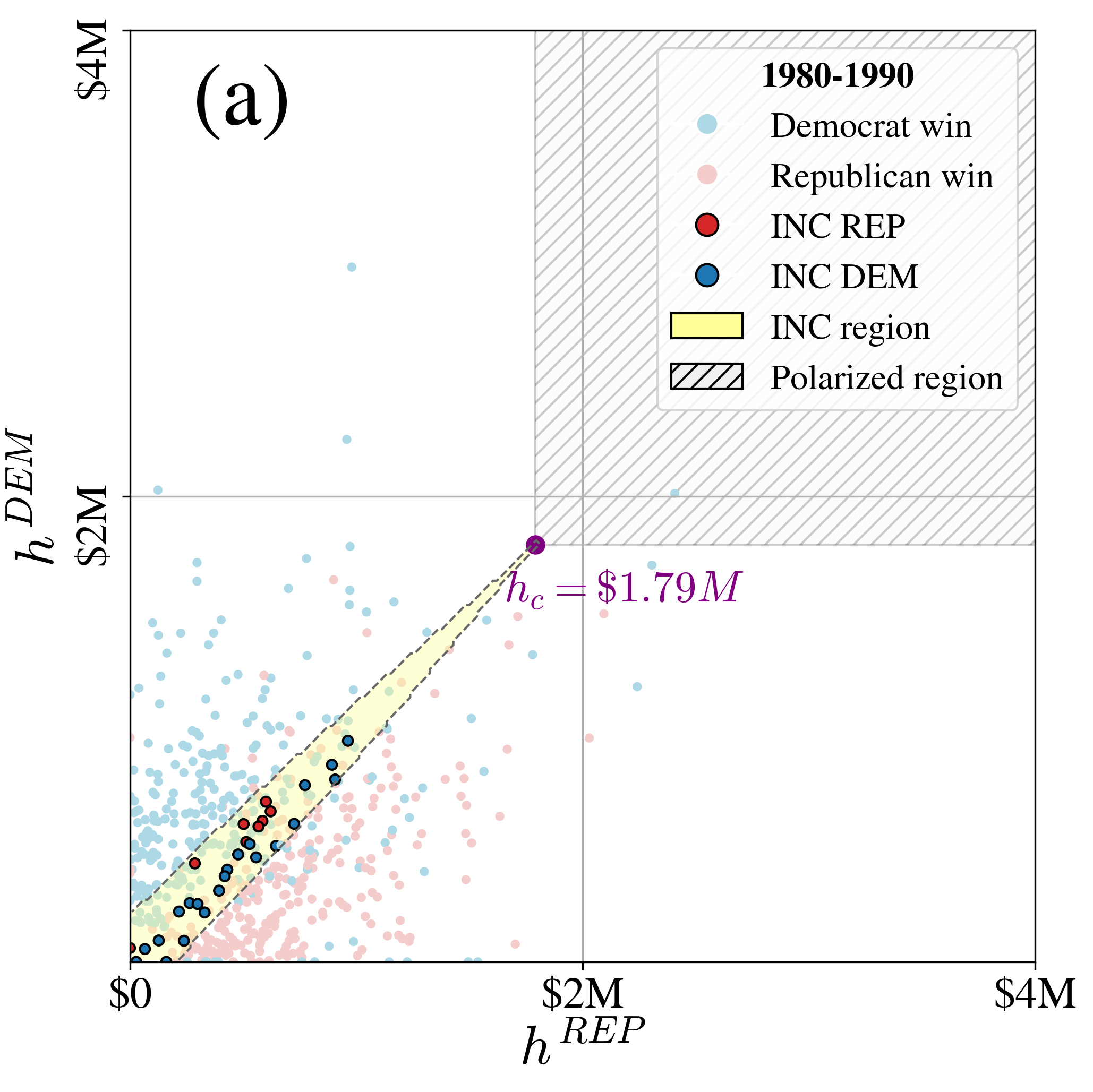}
\includegraphics[width=0.35\linewidth]{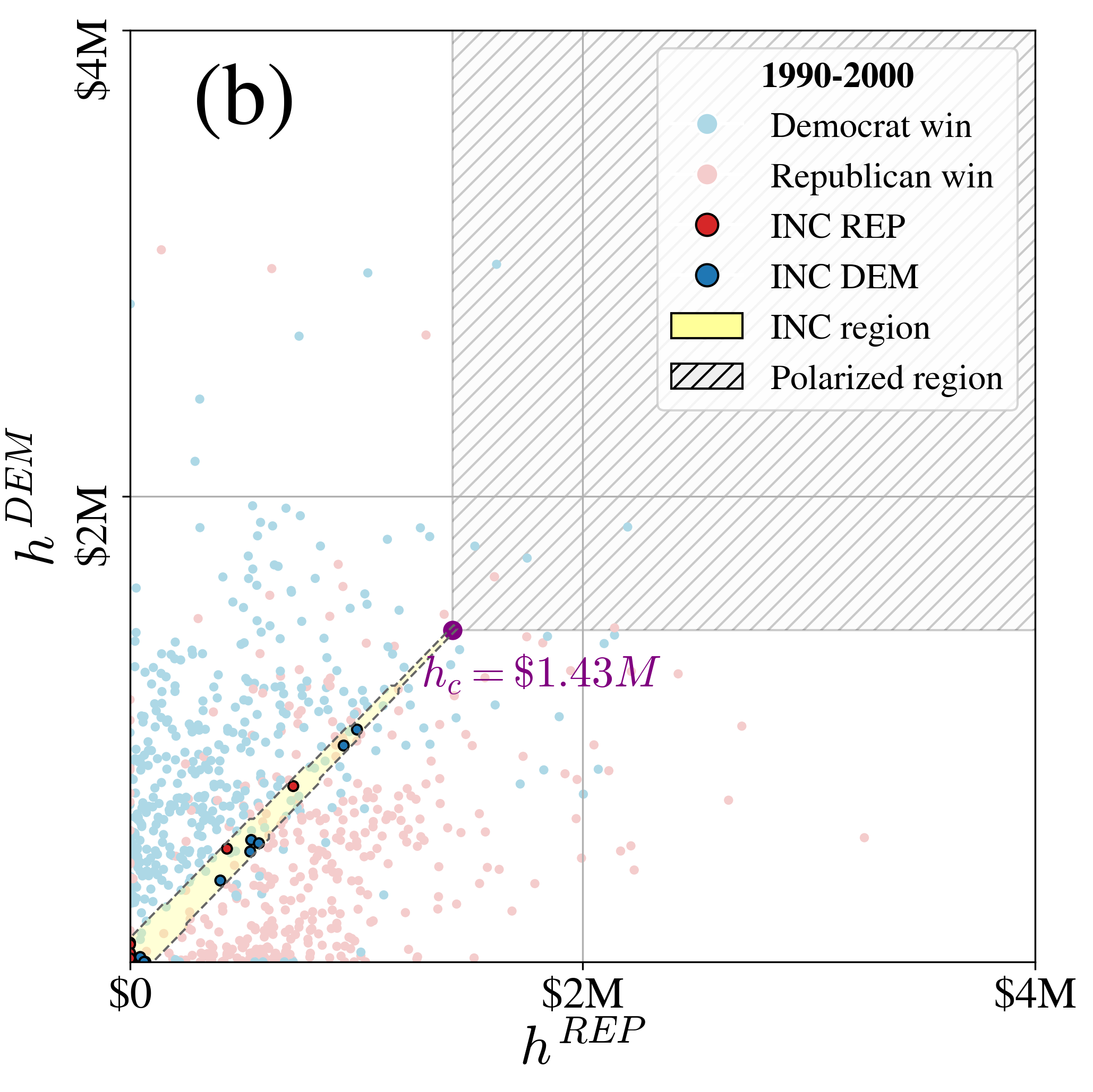}\\
\includegraphics[width=0.35\linewidth]{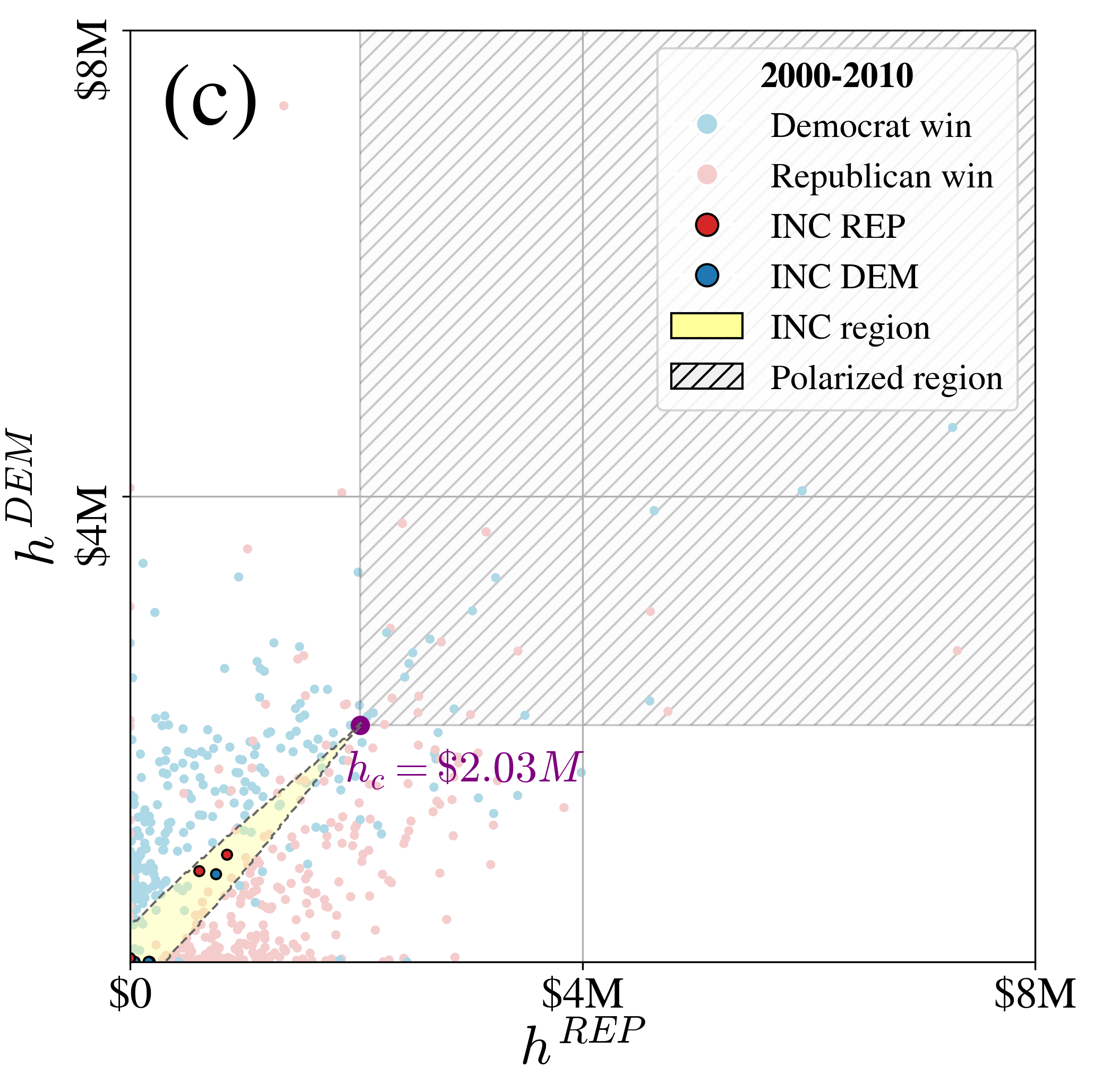}
\includegraphics[width=0.35\linewidth]{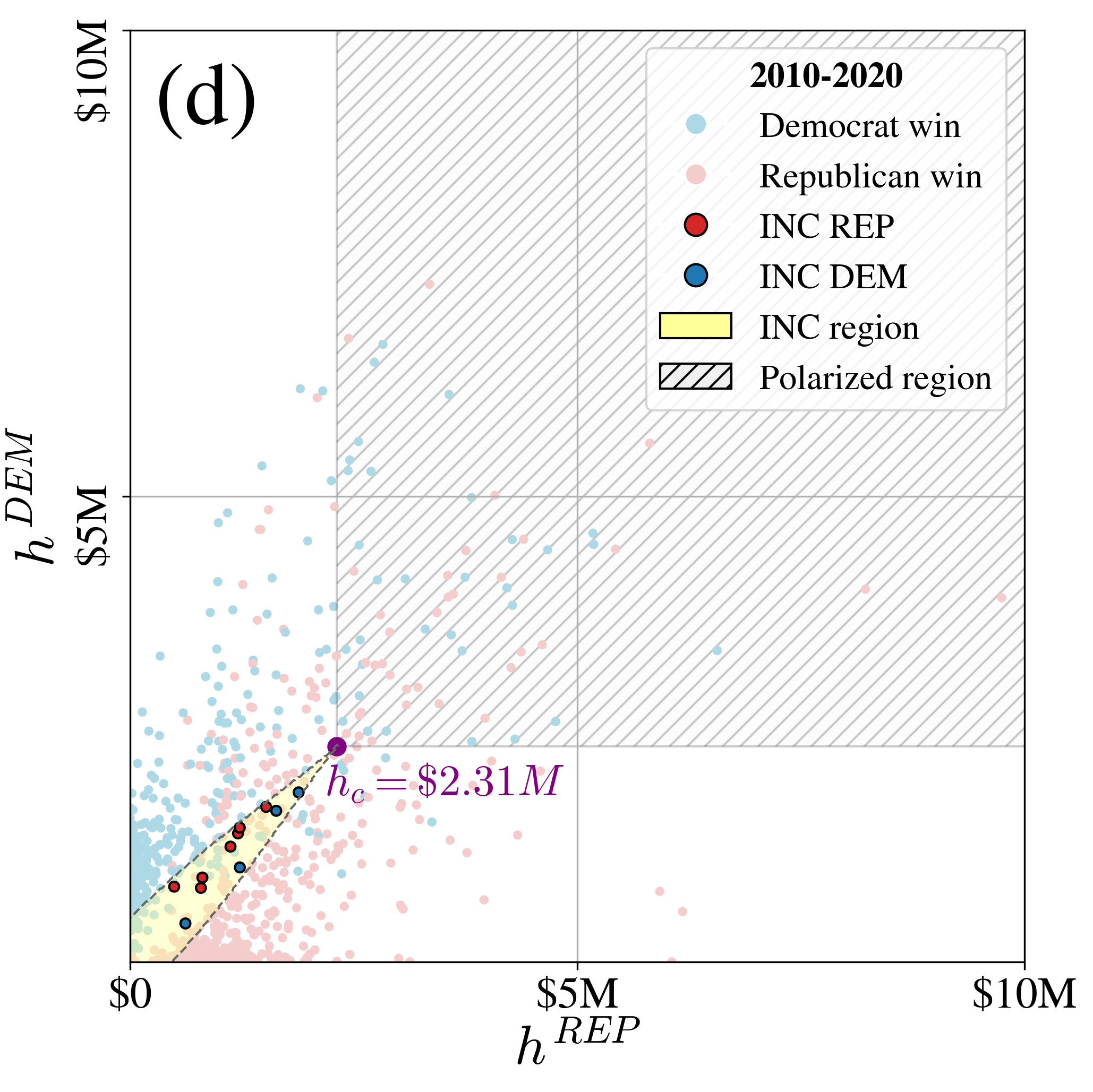}
\caption{Comparison of fitted parameters for US House elections in four decades.}
\label{fig:7}
\end{figure*}

\subsection*{Calibration to US House election data for Republican-leaning races ($p = 0.6 \pm 0.05)$} To illustrate the effectiveness of the classification model also on the case of a subset of races, we choose the races whose previous results were in the range corresponding to  $ p=0.6\pm0.05$. The subset consists of 1145 election races. We find the optimal classification model by finding $T$ and $h_c$ that maximize accuracy. We find out that while the optimal parameters slightly change ($T^\star=0.845$, $h_c=2$ million USD), the overall behavior does not change. We should also stress that since the smaller size of the subset, the statistical tests like the McNemar test exhibit a bit weaker (but still significant) value, which is caused by the fact that the number of points that are classified differently by the optimal model ($T=T^\star$) and the null model ($T=1$) is relatively low.

\begin{figure*}[h!]
\centering
\includegraphics[width=0.35\linewidth]{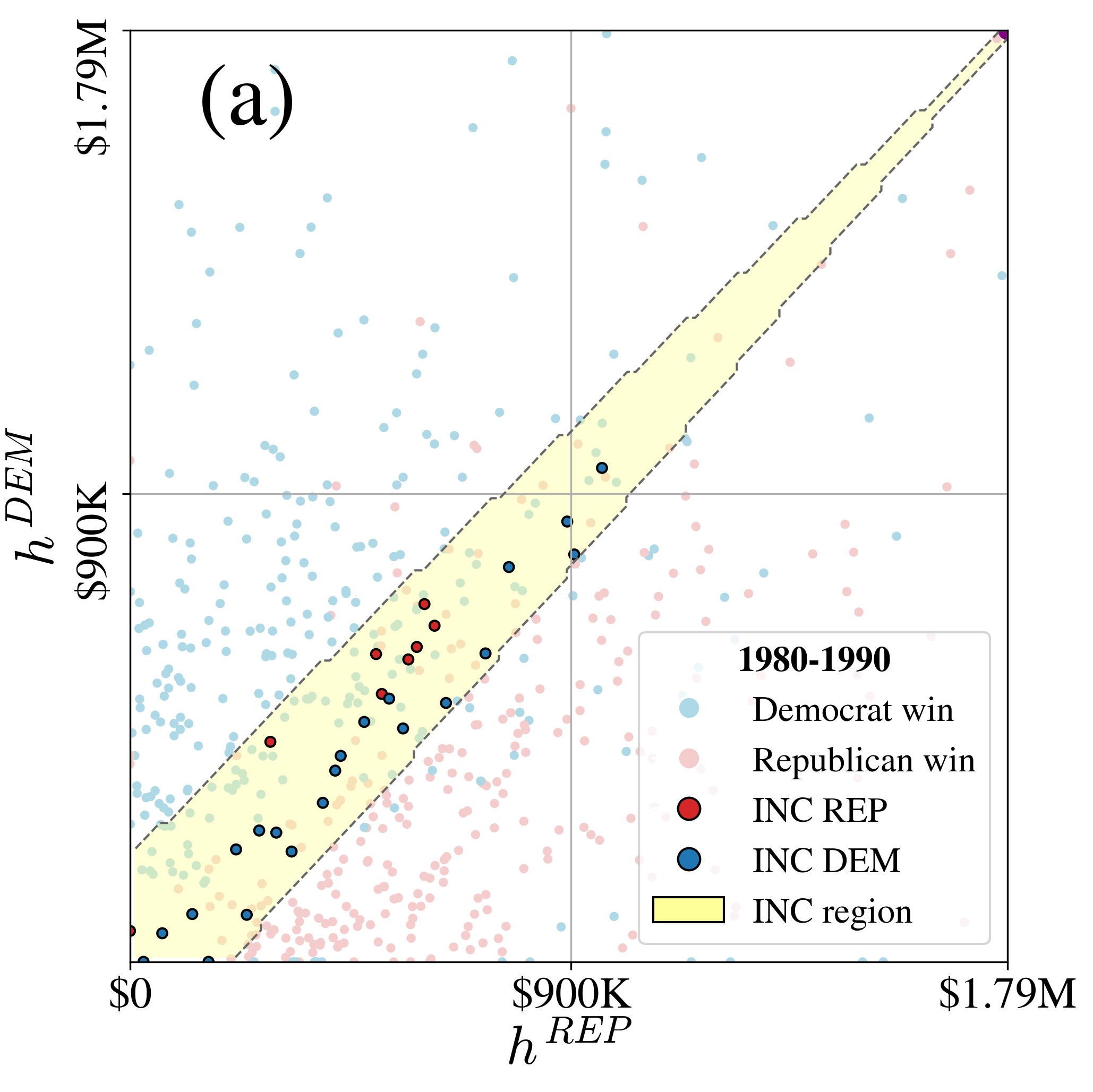}
\includegraphics[width=0.35\linewidth]{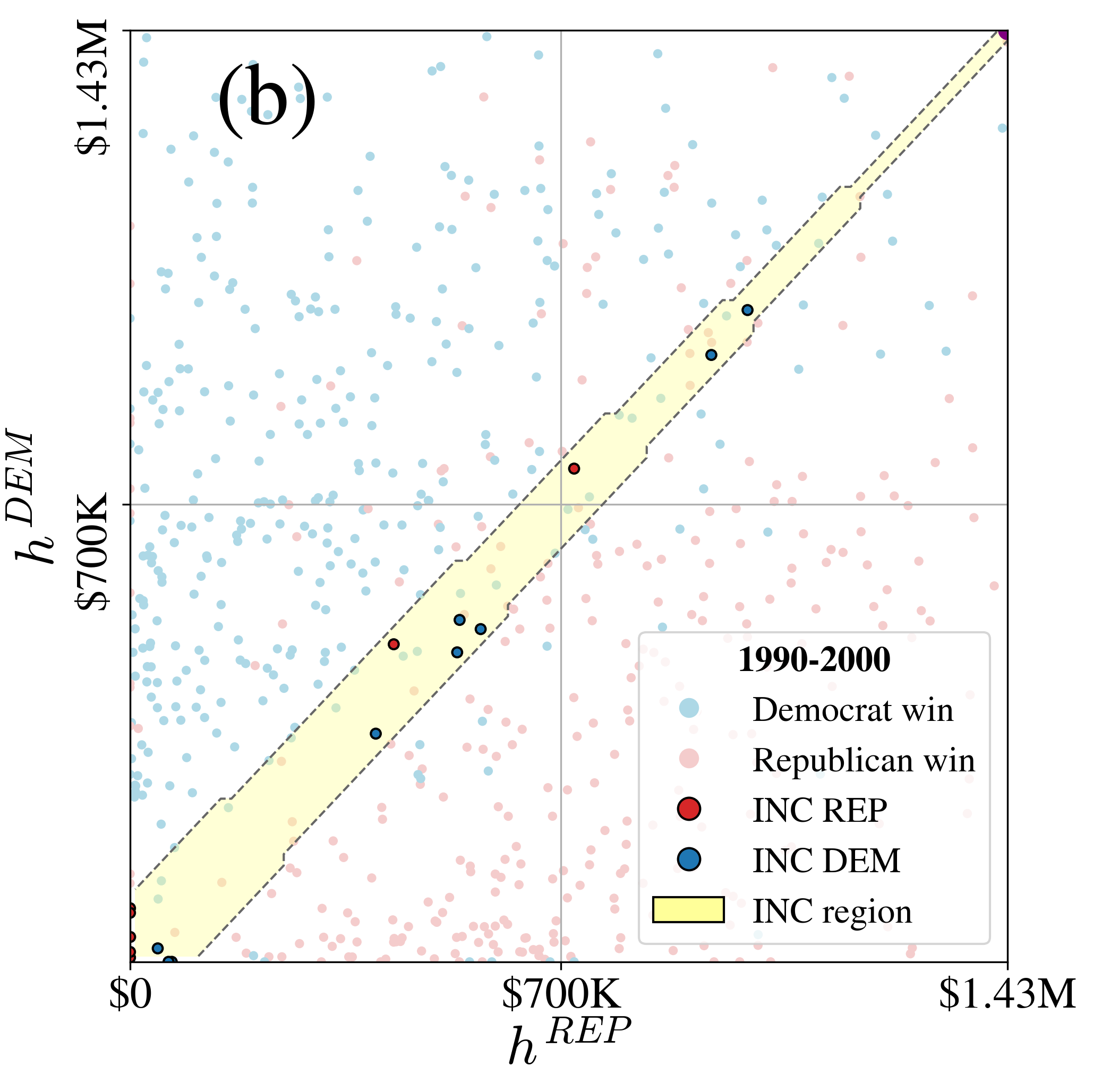}\\
\includegraphics[width=0.35\linewidth]{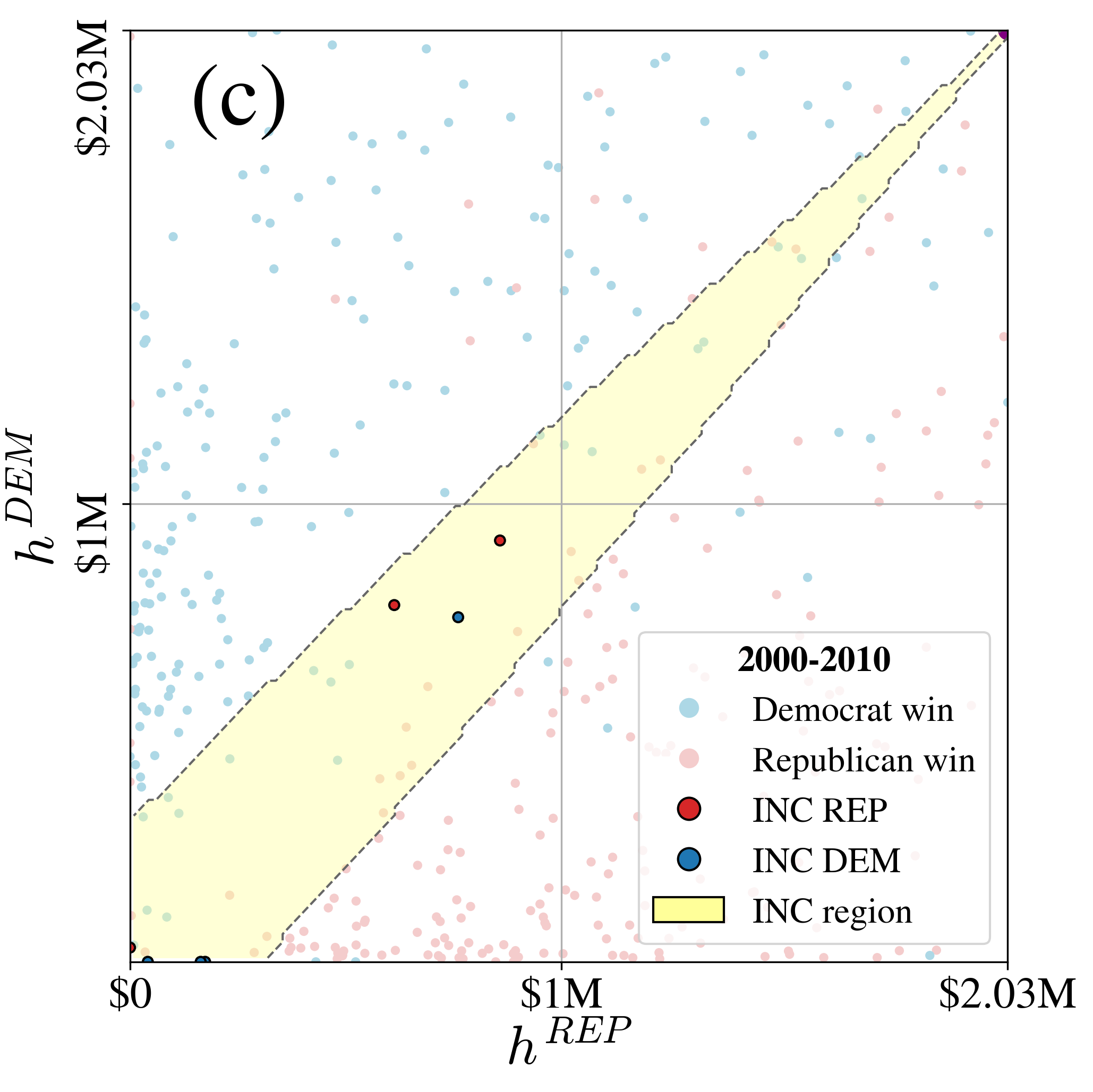}
\includegraphics[width=0.35\linewidth]{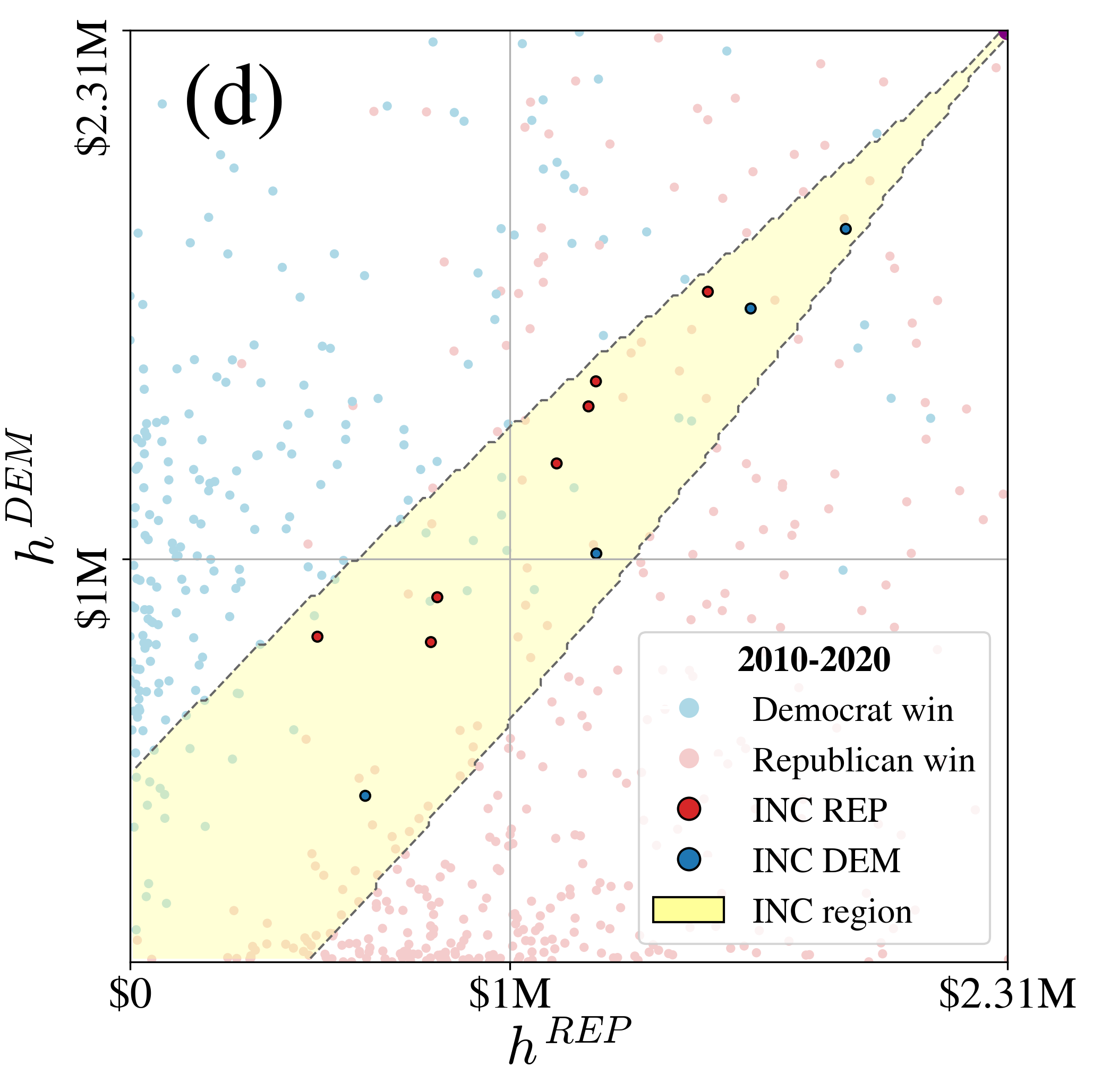}
\caption{Comparison of fitted parameters for US House elections in four decades corresponding to the previous plots, focused on incumbency regions.}
\label{fig:8}
\end{figure*}

\subsection*{Comparison of calibration of US House election data for different time periods} 

To exemplify the robustness of the method and to investigate some aspects of the time-dependence of the thermodynamic quantities, we divide the data into four decades and estimate the model parameters separately. Since the number of data points for the original range of $p$ would be too small, we slightly increase it to $p=0.5 \pm 0.1$. The fits are shown in Fig. \ref{fig:7}, specifically, the incumbency regions are depicted in Fig. \ref{fig:8}; the estimated parameters, together with the accuracy and average spending in the respective period, are summarized in Table \ref{tab:1}. We observe that the temperature is decreasing slightly over time, while the critical spending is increasing. We also observe that the number of incumbents winning despite spending less decreases over time. The accuracy remains almost constant; its slightly smaller value (compared to the values in the main text) is due to the wider region of $p$. 

\begin{table*}[h!]
\begin{tabular}{|c|c|c|c|c|c|}
 \hline 
  period   & \# of races & $T$ & $h_c$ & accuracy & average spending (st.d.)\\
  \hline
 1980-1990 & $581$  & $0.925$  & $\$1.79M$ & $0.8348$ & $\$759K$  ($\$417K$)\\ 
1990-2000 & $743$ & $0.955$ & $\$1.43M$ & $0.8318$& $\$908K$  ($\$501K$)\\
2000-2010 & $515$ & $0.900$ & $\$2.03M$ & $0.8485$ & $\$1.50M$  ($\$971K$)\\
2010-2020 & $765$ & $0.865$ & $\$2.31M$ & $0.8353$ & $\$1.81M$  ($\$1.39M$) \\
\hline
\end{tabular}
\caption{Summary of estimated temperature $T$, critical spending $h_c$ and accuracy for each decade.}
\label{tab:1}
\end{table*}

\begin{figure*}[h]
\includegraphics[width=0.5\linewidth]{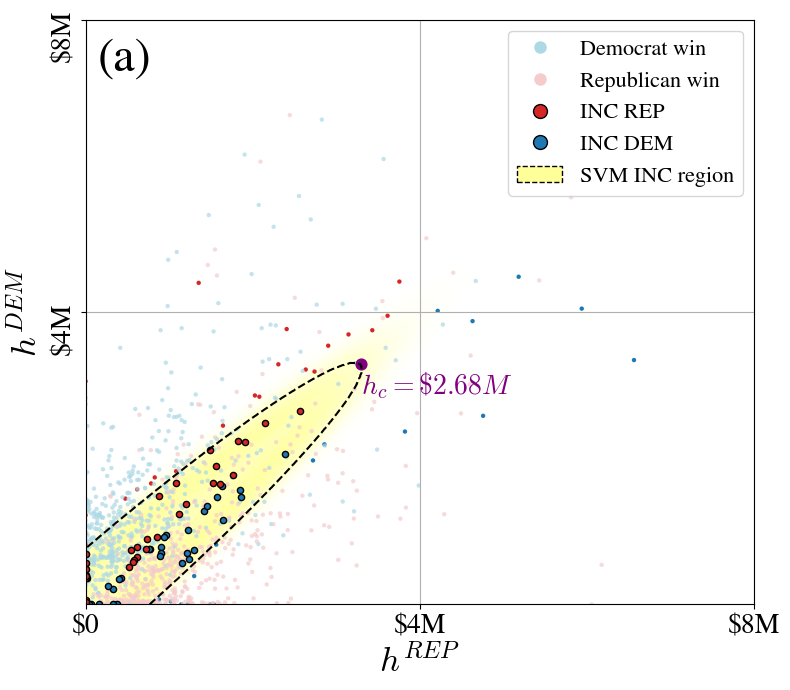}
\caption{Application of SVM to estimate the incumbency region.}
\label{fig:9}
\end{figure*}

\subsection*{Comparison of calibration of US House election data with a support vector machine model} 
Finally, we compare our method with a standard machine learning classification method, particularly the support vector machine. Again, we focus on the close races, i.e., all races with $p=0.5 \pm 0.05$. In order to utilize the natural symmetry of the system, i.e.,  $h^+ = h^-$ leading to $m=0$, we transform the spending data into the following features:
\begin{eqnarray*}
 \sigma_h &=& \frac{{h^+}+{h^-}}{2}\, ,\\
 \delta_h &=& h^{+}-h^{-}\, .
\end{eqnarray*}
We then use the linear SVM on quadratic features,  which is equivalent to the degree-2 polynomial SVM. Additionally, we add a binary feature indicating whether the winner was an incumbent. Since the training data is very unbalanced, we had to upsample the data to approximately equalize the number of instances: more spending by the winner and less spending by the incumbent. The trained accuracy is very high ($94\%$). By transforming the SVM back into the original space using
\begin{eqnarray*}
h^+ &=&  \sigma_h + \delta_h/2\, ,\\
h^- &=& \sigma_h - \delta_h/2\, .
\end{eqnarray*}
We transform the classifier into the original space. The fitted region is depicted in Fig. \ref{fig:9}. By measuring the intersection of the incumbency region with the diagonal ($h^+=h^-$), we obtain the estimate for the equivalent of the critical threshold, which is here $h^*=\$2.68M$. This threshold is higher than predicted by the model in the main text. Furthermore, the accuracy on the whole dataset is lower (only $80\%$), possibly due to overfitting. 
\end{document}